\documentclass[twocolumn,apl,amsmath,amssymb,showpacs,superscriptaddress]{revtex4-1}
\usepackage{epsf}      
\usepackage{graphicx}
\usepackage{color}
\usepackage{soul}
\usepackage{gensymb}
\usepackage{sidecap}
\usepackage{amsmath}
\usepackage{mathtools}

\begin{document}
\title{Probing the electronic and local structure of Sr$_{2-x}$La$_x$CoNbO$_6$ using near-edge and extended x-ray absorption fine structures}
\author{Ajay Kumar}
\affiliation{Department of Physics, Indian Institute of Technology Delhi, Hauz Khas, New Delhi-110016, India}
\author{Rishabh Shukla}
\affiliation{Department of Physics, Indian Institute of Technology Delhi, Hauz Khas, New Delhi-110016, India}
\author{Ravi Kumar}
\affiliation{Atomic and Molecular Physics Division, Bhabha Atomic Research Centre, Mumbai 400085, India}
\author{R. J. Choudhary}
\affiliation{UGC-DAE Consortium for Scientific Research, Indore-452001, Madhya Pradesh, India}
\author{S. N. Jha}
\affiliation{Beamline Development and Application Section, Physics Group, Bhabha Atomic Research Centre, Mumbai 400085, India}
\author{R. S. Dhaka}
\email{rsdhaka@physics.iitd.ac.in}
\affiliation{Department of Physics, Indian Institute of Technology Delhi, Hauz Khas, New Delhi-110016, India}
\date{\today}      

\begin{abstract}
We report the electronic and local structural investigation of double pervoskites Sr$_{2-x}$La$_x$CoNbO$_6$ ($x=$ 0--1) using x-ray absorption near edge structure (XANES) and extended x-ray absorption fine structures (EXAFS) at the Nb, Co, and Sr $K$-edges. The $ab$ $initio$ simulations and detailed analysis of the Nb and Co $K$-edge XANES spectra demonstrate that the observed pre-edge features arise from the transition of 1$s$ electrons to the mixed $p-d$ hybridized states. We reveal a $z$-out Jahn-Teller (JT) distortion in the CoO$_6$ octahedra, which decreases monotonically due to an enhancement in the JT inactive Co$^{2+}$ ions with $x$ . On the other hand, the $z$-in distortion in NbO$_6$ octahedra remains unaltered up to $x=$ 0.4 and then decreases with further increase in $x$. This sudden change in the local coordination around Nb atoms is found to be responsible for the evolution of the antiferromagnetic interactions in $x \geqslant$ 0.6 samples. Also, we establish a correlation between the degree of octahedral distortion and intensity of the white line feature in the XANES spectra and possible reason for this are discussed. More interestingly, we observe the signature of KN$_1$ double electron excitation in the Sr $K$-edge EXAFS spectra for all the samples, which is found to be in good agreement with the $Z$+1 approximation. Further, the Co L$_{2,3}$ edge shows the reduction in the crystal field strength and hence an increase in the charge transfer energy ($\Delta_{ct}$) with the La substitution. 
\end{abstract}

\maketitle
\section{\noindent ~Introduction}
The octahedrally-coordinated Co possesses the intriguing magnetic, transport, optical, and electronic properties due to the small energy difference between its various oxidation and spin-states \cite{Mabbs_book_73, Lloret_ICA_08, Liu_PRB_18}. The strength of the crystal field (CF) is a crucial measure to govern these properties, as a delicate competition between the Hund's exchange energy and CF splitting decide the electronic distribution in the CF splitted t$_{2g}$ and e$_g$ states \cite{Raccah_PR_67, Vanko_PRB_06, Merz_PRB_11}. In this case, the Co--O bond distance plays the key role in controlling the CF strength. For example, Chen {\it et al.} have reported a correlation between the Co--O bond length and spin-state transition of Co$^{3+}$ from high spin (HS) state (3$d^6$; t$_{2g}^4$e$_g^2$) to the low spin (LS) state (3$d^6$; t$_{2g}^6$e$_g^0$) in SrCo$_{0.5}$Ru$_{0.5}$O$_{3-\delta}$ using the pressure (hydrostatic) dependent x-ray absorption and emission spectroscopy \cite{Chen_JACS_14}. It was found that a Co--O bond length $>$1.93 \AA $\space$ favors the HS sate, whereas a smaller value results in the most stable LS state due to the large CF splitting in the latter \cite{Chen_JACS_14}. Also, an increase in the Co--O--Co bond angle enhance the Co $3d$--O $2p$ orbital hybridization, which stabilize the intermediate spin (IS) (3$d^6$; t$_{2g}^5$e$_g^1$) state of Co in LaCoO$_3$ \cite{Knizek_PRB_05}. On the other hand, the Co$^{2+}$ usually stabilizes in the HS state (3$d^7$; t$_{2g}^5$e$_g^2$) due to the weaker CF as compared to Co$^{3+}$ in Co based perovskite oxides \cite{Mabbs_book_73, Lloret_ICA_08, Kumar_PRB1_20}. Thus a precise understanding of the chemical and local coordination environment around Co and its evolution with the external parameters like temperature, mechanical pressure, and chemical pressure (doping) and hence CF strength is extremely important to engineer the various  physical properties of the complex oxides and their vast use in the technological applications \cite{Tao_NM_03, Chakrabartty_NP_18}. 

In this context, a stable crystal structure of the perovskite oxides [ABO$_3$; A: rare earth/alkali earth metals, B/B: transition metals (TM)] provides a suitable matrix to accommodate the wide range of elements with tuneable CF strength and hence degree of TM--O orbital hybridization \cite{Medarde_PRB_06, Krayzman_PRB_06, Shukla_PRB_18}. Further, a 50\% substitution of B-site cations results in the alternating ordering of BO$_6$ and B$^\prime$(say)O$_6$ octahedra, where the degree of ordering depends on the ionic and valence mismatch between two $B$-site cations \cite{Anderson_SSC_93, King_JMC_10,Vasala_SSC_15, Galasso_JPC_62}. This ordering results in the additional B-O-B$^\prime$-O-B exchange interactions along with the conventional B-O-B channels, resulting in the exotic magnetic and transport properties \cite{Niebieskikwiat_PRB_04, Sanchez_PRB_02, Jung_PRB_07, Erten_PRL_11, Frontera_PRB_04}. For example, the anti-site disorder driven several exchange interactions are found to play a crucial role in determining the inverse exchange bias effect observed in Gd$_2$CoRuO$_6$ \cite{Das_PRB_20}. In the double perovskite family, the Sr$_2$CoNbO$_6$ is one of the most interesting candidates, where high resolution electron microscopy (HREM) show the $B$-site ordered domains of less than 100 \AA $\space$, however no such evidence was observed in the x-ray and neutron powder diffraction measurements due to their large coherence length as compared to the size of the ordered domains \cite{Kumar_PRB1_20, Azcondo_DT_15}. This is because Sr$_2$CoNbO$_6$ is a border line candidate of  the ordered-disordered configurations due to the moderate ionic and valence mismatch between Co$^{3+}$ and Nb$^{5+}$ ions and even a small perturbation can transform it into either regime \cite{King_JMC_10,Vasala_SSC_15}. This intriguing crystal structure of Sr$_2$CoNbO$_6$ is closely related to its observed colossal dielectric properties, complex ac impedance spectroscopy, and cluster glass like magnetic ground state \cite{Wang_AIP_13, Bashir_SSS_11, Kumar_PRB2_20}. 

Recently, we studied the doping induced spin-state transition of Co$^{3+}$ from HS to LS/IS states in Sr$_{2-x}$La$_x$CoNbO$_6$ ($x= $0--1) samples using the magnetization measurements \cite {Kumar_PRB1_20}. Here it is important to note that the substitution of each La$^{3+}$ ion at Sr$^{2+}$ site transform one Co from 3+ to 2+ state, which systematically reduces the CF strength and stabilize the Co$^{2+}$ in HS state \cite{Mabbs_book_73, Kumar_PRB1_20, Bos_PRB_04}. A very small tetragonal ($I4/m$) distortion was observed in the $x=$ 0 sample from the perfect cubic ($Pm\bar3m$) symmetry, which increases with the La substitution ($x$) up to the $x=$ 0.4, and then transformed into the monoclinic ($P2_1/n$) structure for $x\geqslant$0.6 samples \cite{Kumar_PRB1_20}. Further, an increase in the valence mismatch between $B$-site cations ($\Delta$V= 2 $\rightarrow$ 3) with $x$ transform the disordered parent sample to the highly ordered state in the $x=$ 1 sample \cite{Kumar_PRB1_20}. Importantly, an abrupt enhancement in the $B$-site ordering and evolution of antiferromagnetic (AFM) interactions were reported for $x\geqslant$0.6 samples \cite{Kumar_PRB1_20, Kumar_PRB2_20}, which indicate the strong correlation between the local crystal structure and the complex magnetic interactions present in these samples. Also, the specific heat measurements indicate the persistence of the discrete energy states resulting from the crystal field splitting, spin-orbit coupling, and octahedral distortion \cite{Kumar_PRB2_20}. However, extent of the octahedral distortion, which can play a key role in governing the various spin-states of Co$^{3+}$, and its evolution with the La substitution have not been quantified. Therefore, a systematic study of change in the electronic structure and local coordination around different cations is essential to understand the origin of complex magnetic interactions in these samples. In this direction, x-ray absorption spectroscopy (XAS) is a versatile tool to investigate the ligand-field, spin-state, valence state, centrosymmetry, metal-ligand overlap, and the local coordination geometry \cite{Bunker_book_10}. 

Therefore, in this paper, we use x-ray absorption spectroscopy to investigate the element specific electronic and local structure of Sr$_{2-x}$La$_x$CoNbO$_6$ ($x=$ 0--1), and focus on the Nb, Co and Sr $K$-edges in both the XANES and EXAFS regions. The Nb $K$-edge XANES shows a very weak pre-edge feature in spite of the completely unoccupied 4$d$ states, indicating the presence of Nb atoms in the centrosymmetric octahedral environment. However, a strong pre-edge feature is observed in case of Co $K$-edge XANES spectra, which indicates its off-center displacement in the CoO$_6$ octahedra. The analysis suggests that pre-edge features in both Nb and Co $K$-edge spectra result from the transition of 1$s$ electrons to the mixed $p-d$ hybridized states. We demonstrate that the intensity and position of the post edge feature in the Nb $K$-edge spectra depend on the nature and arrangement of the $A$-site atoms, respectively. Moreover, we found a strong correlation between the intensity of the white line and degree of octahedral distortion in CoO$_6$ and NbO$_6$, where a higher degree of octahedral distortion (compression as well as elongation) results in the reduction of white line intensity of the respective $K$-edge absorption spectra. The Co $K$-edge spectra show the monotonic reduction in the distortion in CoO$_6$ octahedra with the La substitution ($x$). However, the Nb $K$-edge spectra show an abrupt change in the NbO$_6$ octahedral distortion for $x\geqslant$0.6 samples, causing a sudden enhancement in the degree of $B$-site ordering, resulting in the observed AFM interactions at this critical doping. Further, the  presence of multielectronic excitation in all the samples in Sr $K$-edge EXAFS spectra is confirmed using Z+1 approximation as well as La $L_3$-edge absorption spectra. The Co L$_{2,3}$ edge spectra show the presence of Co$^{3+}$ in IS/HS state and significant reduction in the crystal field strength with $x$. 

\section{\noindent ~Experimental}

Polycrystalline samples of Sr$_{2-x}$La$_{x}$CoNbO$_{6}$ ($x=$ 0--1) were synthesized by usual solid-state route, more details and characterization can be found in ref.~\cite{Kumar_PRB1_20}. The x-ray absorption spectroscopic data were recorded at Nb, Co and Sr $K$-edges in the transmission mode for all the samples using BL-09 (4--25~keV) beam line at Indus-2 synchrotron source (2.5~GeV, 200~mA) at Raja Ramanna Center for Advanced Technology (RRCAT) in Indore, India. The polycrystalline powdered samples were thoroughly mixed in the boron nitride (BN) using a mortar pestle in such a proportion that absorption edge jump ($\Delta\mu$) lies between 0.7--1, calculated using the XAFSmass code \cite{Klementiev_JPCS_16} and then pressed the total 100~mg powder into the circular disc of $\sim$15 mm diameter, and sandwich between the kapton tape for the stability. A Si(111) double crystal monochromator was used for the precise energy selection of the incident beam with a resolving power of $\Delta$E$/E\approx$10$^{-4}$. More technical details of the beamline can be found in  \cite{Basu_JPCS_14}. We calibrate all the recorded spectra using the following energy dependent relation \cite{Bunker_book_10}:
\begin{equation}
E_{\rm calb}= \frac{gE_{\rm expt}}{\sqrt{(E^2_{\rm expt}-g^2)}sin(\Delta \theta)+gcos(\Delta \theta)}, 
\label{calb}
\end{equation}
where, $E_{\rm calb}$ and $E_{\rm expt}$ are the calibrated and experimentally recorded energies, respectively, $g\approx$1977.1~eV for the Si(111) crystal, and $\Delta \theta$ was calculated, by recording the same absorption edge for the reference metal foils, using the equation below:
\begin{equation}
\Delta \theta=sin^{-1}\left( \frac{g}{E_0}\right)-sin^{-1}\left( \frac{g}{E_{\rm calc}}\right), 
\label{calb2}
\end{equation}
where, $E_{\rm calc}$ is the position of the first maxima of the first derivative of the absorption coefficient of the reference metal foil and E$_0$ is its known reference energy taken at 7709 and 18986~eV for Co and Nb $K$-edges, respectively from ref.~\cite{Bunker_book_10}. Due to unavailability of the reference metal foil, Sr K-edge spectra was calibrated by aligning the first maxima of the first derivative for the $x=$ 0 sample at 16105~eV (for Sr$^{2+}$ \cite{Sahai_JCIS_2000}) using the above energy dependent calibration equations (\ref{calb}, \ref{calb2}) and then other samples were calibrated with reference to $x=$ 0.

 A linear pre-edge and a spline post edge background were subtracted from the XAS spectra and the resultant spectra were normalized at 800~eV above the edge jump for all the samples. We use the Gaussian peak shape and an error step function (accounting for the transition to the continuum) to fit the XANES region (up to $\sim$50~eV above the edge jump) in the athena software \cite{Ravel_JSR_05}. The position and FWHM of the step function have been decided in such a way that it touches the rising edge region with the similar slope as that of edge and height was adjusted such that it touches the lowest feature of the XANES spectra under consideration. The FWHM and height of the error function were fixed for all the samples, whereas its position was varied as per the change in the position of rising edge region due to the valence shift. The FEFF9.6.4 code for the $ab$ $initio$ calculations, using the full multiple scattering within a sphere around the absorbing atom with the muffin tin approximation in a self-consistent loop \cite{Rehr_PCCP_2010, Rehr_RMP_2000}, is used to simulate the XANES spectra of Nb, Co, and Sr $K$-edges. The input structural parameters are taken from the refinement of diffraction patterns where the Nb and Co atoms are considered in completely ordered and disordered states at two different B-sites for the $x=$ 0 and 1 samples, respectively \cite{Kumar_PRB1_20, Bos_PRB_04}. We use Hedin-Lundqvist (HL) exchnage correlation potential to simulate Nb and Co $K$-edges, and Dirac-Hara+HL for Sr $K$-edge XANES spectra with a fully screened core-hole (final state rule).

The Co $L_{2,3}$-edge XAS spectra were recorded in the total electron yield mode at beamline BL-01of the same synchrotron source. A constant energy shift in each spectra was employed to align the center of L$_3$ edge of the $x=$ 1 sample corresponding to the octahedrally coordinated Co$^{2+}$ in HS state, as in ref.~\cite{Merz_PRB_11}, due to the unavailability of the reference sample. The Co $L_{2,3}$-edge is simulated by the CTM4XAS software based on the atomic-multiplet, crystal-field, and charge transfer approach \cite{Groot_PRB_90, Stavitski_Micron_10}, as single electron approach is unable to reproduce the transition metal $L$-edge XAS spectra. The simulated spectra of HS Co$^{2+}$ and HS Co$^{3+}$ were arbitrarily shifted to align with the $x=$ 1 and 0.2 samples, respectively.  

\begin{figure*} 
\includegraphics[width=7.1in]{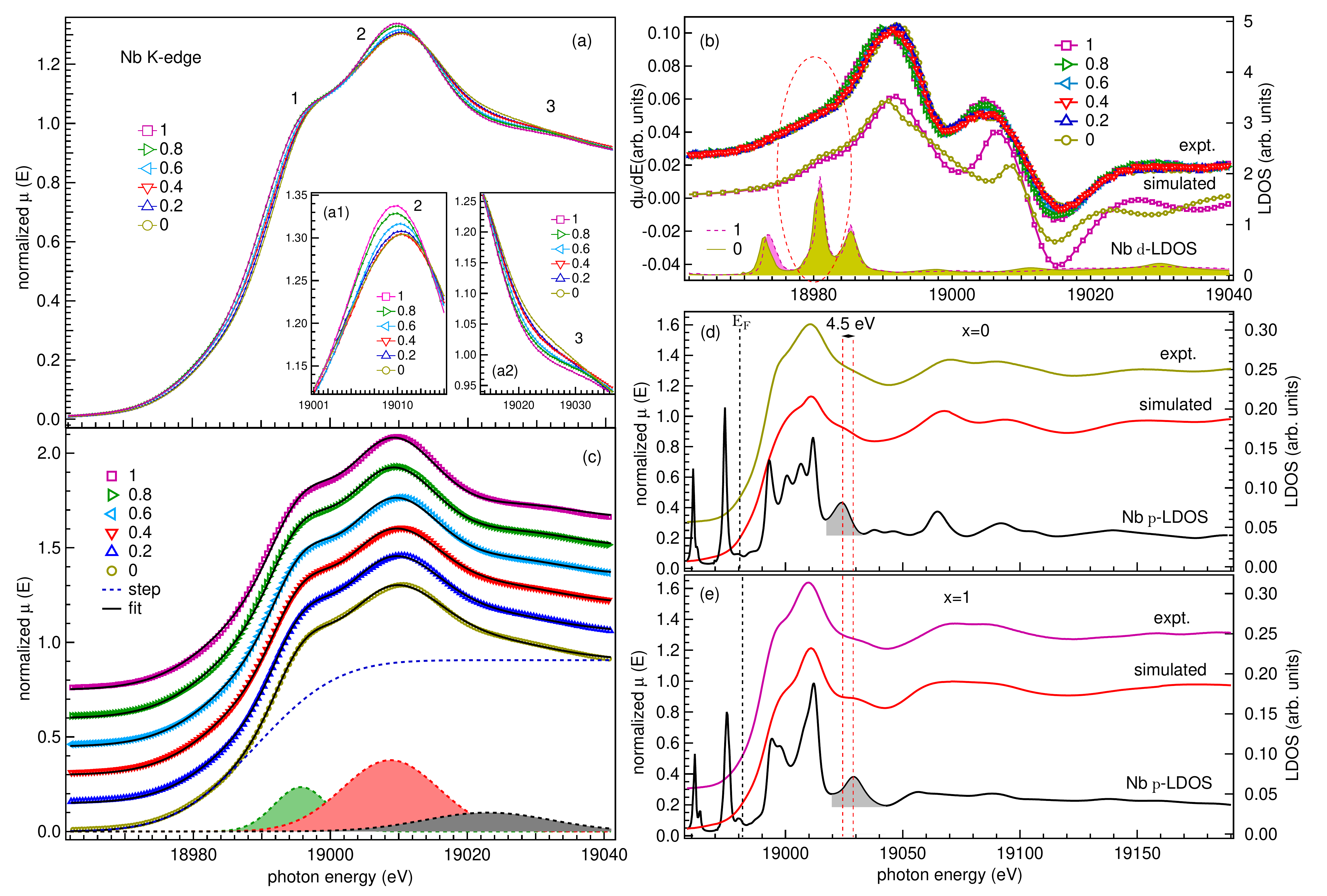}
\caption {(a) The normalized Nb $K$-edge XANES spectra of Sr$_{2-x}$La$_x$CoNbO$_6$ ($x=$ 0--1) samples where 1, 2, 3 represent the main absorption features, and enlarged view of features 2 and 3 are shown in the insets. (b) The first derivative of the experimental absorption spectra of the Nb $K$-edge along the simulated spectrum for the $x=$ 0 and 1 samples and their corresponding Nb$_{\rm abs}$ $d$-LDOS in the XANES region. The red dashed ellipse highlight the presence of pre-edge feature in both experimental and simulated spectra, and sharp Nb$_{\rm abs}$ $d$-LDOS at this energy. All the experimental spectra have been vertically shifted by fix 0.025 units for the clarity. (c) The best fit of XANES spectra using the Gaussian peak shape for all the samples, and an error function accounting for the different absorption features and the edge jump, respectively, shown for the $x=$ 0 sample. Here, each spectrum is vertically shifted cumulatively by 0.15 units for the clear presentation. (d, e) The experimental and simulated Nb $K$-edge absorption spectra along the calculated $p$-LDOS of absorbing Nb atoms for $x=$ 0 and 1 samples. The experimental spectra have been vertically shifted by 0.3 units for the clarity. The black dashed line represent the Fermi level and red dashed lines indicate the shift in the Nb$_{\rm abs}$ $p$-LDOS peak corresponding to the feature 3 (shaded region).} 
\label{XANES_Nb_K}
\end{figure*}

\section{\noindent ~Results and discussion}

\subsection{\noindent ~X-ray absorption near edge structure:}

The normalized XANES spectra of Nb $K$-edge is presented in Fig.~\ref{XANES_Nb_K}(a) for Sr$_{2-x}$La$_x$CoNbO$_6$ ($x=$ 0--1) samples to probe the electronic structure and multiple scattering (MS) mechanism around the Nb atoms. We first discuss the pre-edge region, where no significant intensity is evident around 18980~eV at the first glance, in spite of the completely unoccupied Nb 4$d$ states (4$d^0$ for Nb$^{5+}$) in these samples. Note that the low intense pre-edge features in the XANES spectra of TM oxides result from the transition of 1$s$ electrons to the unoccupied pure $d$ states (quadruple transitions; $\Delta l= $2), and/or transition into the $d$ states hybridized either with their $p$ states or oxygen 2$p$ states \cite{Joly_PRL_99, Yamamoto_XRS_08, Farges_PRB_97, Krayzman_PRB_06}. The probability of the pure quadruple transitions is significantly lower and hence intensity of the pre-edge feature(s) depends mainly on the degree of $p-d$ hybridization \cite{Farges_PRB_97, Yamamoto_XRS_08}. The group theory predict the absence of the $p-d$ mixing and hence the pre-edge feature in case of the ideal octahedral symmetry ($O_h$), i.e., when absorbing atom lies in the centrosymmetric $O_h$ environment and intensity of the pre-edge features increases with increase in the off-center displacement of the absorbing atoms \cite{Yamamoto_XRS_08, Vedrinskii_JPCM_98, Frenkel_PRB_04}. Interestingly, we observe an asymmetricity in the first derivative of the absorption coefficient in the pre-edge region for all the samples as highlighted by the dotted ellipse in Fig.~\ref{XANES_Nb_K}(b), which indicates the minimal off-center displacement of the Nb atoms in NbO$_6$ octahedra. However, the weak pre-edge feature can also result from the thermal assisted dynamic distortion in the NbO$_6$ at room temperature, as reported even in centrosymmetric perovskite EuTiO$_3$ \cite{Ravel_Ferroelectrics_98}. In order to understand the origin of weak pre-edge feature, the Nb $K$-edge XANES spectra have been simulated for the $x=$ 0 and 1 samples using the FEFF 9.6.4 code for a spherical cluster of 45 atoms. Here, it is interesting to note that we have allowed only dipole transitions in the simulations and the weak pre-edge feature is well reproduced in the simulated spectra for both the samples, as compared in Fig.~\ref{XANES_Nb_K}(b). In fact, we find that there is no significant  enhancement in the pre-edge feature by introducing the quadruple transitions from 1$s$ to the unoccupied $d$ states (see Fig.~1 of \cite{Kumar_XAS_SI}). This suggest that the pre-edge feature in all the samples is predominantly due to the dipolar transition from the 1$s$ to the $d$ states of Nb that hybridize with its $p$ states and/or $p$-states of oxygen. To understand further, we plot the $d$-local density of states (LDOS) for the absorbing Nb atoms (Nb$_{\rm abs}$) in Fig.~\ref{XANES_Nb_K}(b) for the $x=$ 0 and 1 samples. Interestingly, we observe a sharp Nb$_{\rm abs}$ $d$-LDOS at around 18980~eV corresponding to the weak pre-edge feature, which suggests the key role of the Nb $d$ states in determining the pre-edge feature in these samples. However, a slight discrepancy in intensity behavior between Nb $d$-LDOS and the simulated spectra is evident in the pre-edge region for the $x=$ 0 and 1 sample. The possible reason can be a change in the degree of $p$--$d$ mixing due to different octahedral distortion in two cases, which is also observed in the EXAFS analysis, discussed in the next subsection. On the other hand, this small change in the pre-edge feature with $x$ is difficult to detect in the experimental spectra.

We now discuss the rising edge (around 18990~eV) and other features (denoted by 1, 2, 3) observed in the Nb $K$-edge, see Fig.~\ref{XANES_Nb_K}(a). We note that there is no significant shift in the rising edge position suggesting the invariance in the valence state of Nb with the La substitution. A comparison with the reference Nb$_2$O$_5$ sample confirm the pentavalent state of Nb in all the samples, see Figs.~2(a, b) of \cite{Kumar_XAS_SI}. Further, two prominent features 1 and 2, which are separated by $\sim$ 13~eV, can be attributed to the 1$s\rightarrow$5$p$ transition and the resonance due to multiple scattering of the photoelectrons from the neighboring atoms, respectively \cite{Ruckman_PRB_98, Lytle_PRB_88, Marini_JPCM_16}. It is interesting to note that the intensity of the feature 2 remains almost invariant up to the $x=$ 0.4 and then increases monotonically with further increase in the La substitution for the $x\geqslant$0.6 samples [see inset (a1) of Fig.~\ref{XANES_Nb_K}(a)]. This is possibly due to an abrupt change in the local coordination environment around the Nb atoms, which can be speculated as the origin of the evolution of antiferromagnetic interactions in the $x\geqslant$0.6 samples \cite{Kumar_PRB1_20}. We will get back to this point during the discussion of the EXAFS part in the next subsection. Also, a post-edge feature (marked as 3) is observed in all the samples, which shifts towards the higher energy side (see Fig.~3 of \cite{Kumar_XAS_SI} for clarity) with a monotonic reduction in its strength [see inset (a2) of Fig.~\ref{XANES_Nb_K}(a)] with the La substitution ($x$). 

To understand the quantitative change in the different spectral features with $x$, we perform the deconvolution of all the spectra using three Gaussian components and a steplike error function, accounting for the continuum excitations, as presented in Fig.~\ref{XANES_Nb_K}(c) and the best fit parameters are summarized in Table I of \cite{Kumar_XAS_SI}. In order to understand the origin of the feature 3, in Figs.~\ref{XANES_Nb_K}(d, e), we show the simulated Nb $K$-edge spectra, which are in good agreement with the experimental spectra in the entire energy range. In fact, the tendency of a doublet around 19050--19110~eV turning into the singlet with the La substitution (highlighted by downward arrows in the inset of Fig.~3 of \cite{Kumar_XAS_SI}) is well reproduced in the simulated spectra, discussed in next paragraph. Note that we use a shift of --22.0~eV and an additional broadening of 2.0~eV in the simulation to account for the uncertainty in the Fermi level and core-hole life time, respectively. Interestingly, we find that the feature 3 shifts towards the higher energy with $x$ and to understand this, we plot the Nb$_{\rm abs}$ $p$-LDOS in Figs.~\ref{XANES_Nb_K}(d, e) for the $x=$ 0 and 1 samples, respectively. A clear shift of around 4.5~eV in the unoccupied density of Nb $p$ states correspond to feature 3 (shaded areas) is observed with $x$, as highlighted by the red vertically dashed lines, which indicate an increase in the probability of 1$s \rightarrow p$ transition at the higher energy in case of the $x=$ 1 sample. This perturbation in the unoccupied Nb $p$-LDOS states with La substitution is originating either from the atomic-like electronic redistribution due to evolution of the new La states or/and from change in the multiple scattering mechanism due to variation in the local coordination environment around the Nb atoms. However, these need to be further explored for complete understanding. 

\begin{figure}
\includegraphics[width=3.45in]{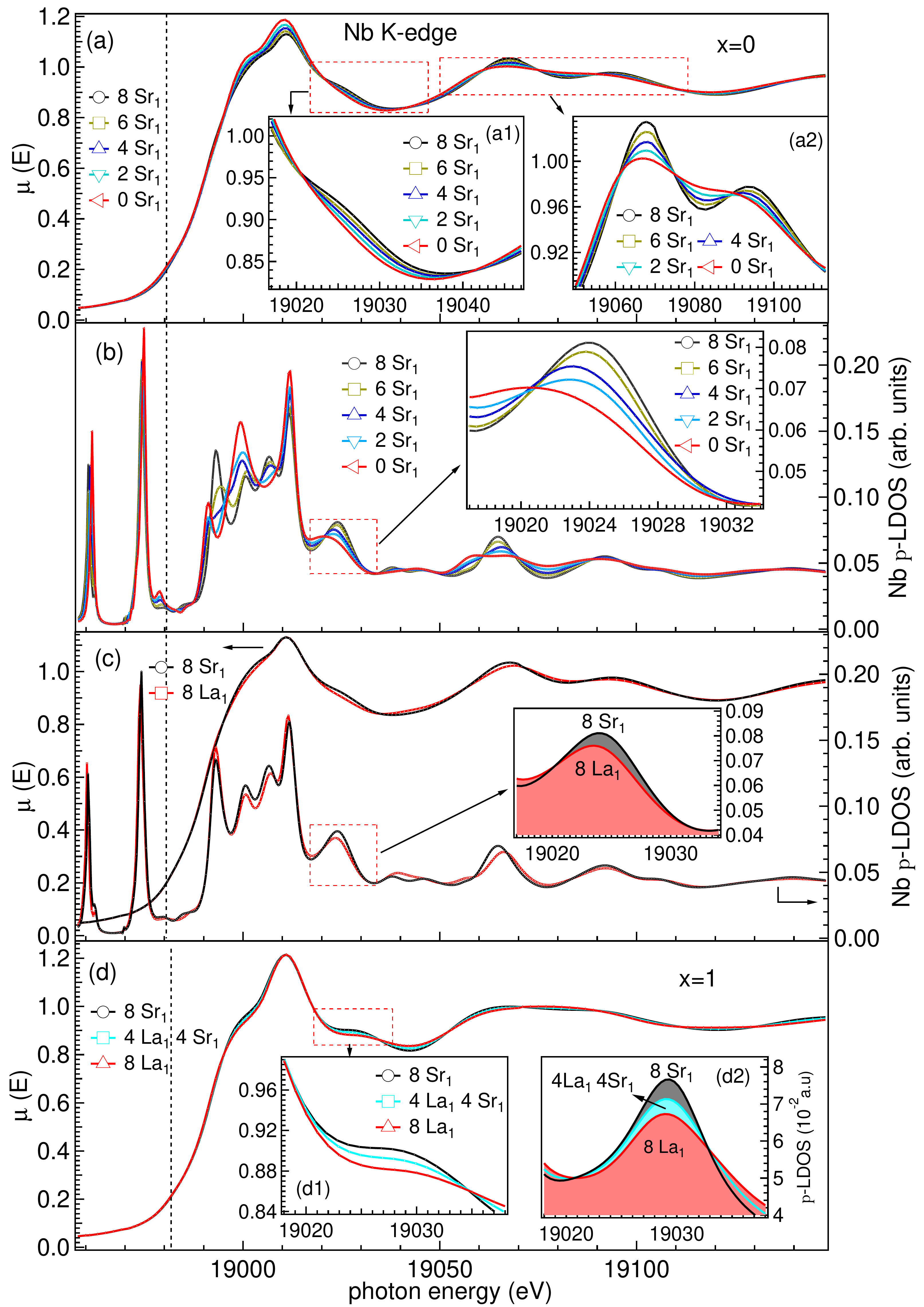}
\caption {(a) The simulated Nb $K$-edge XANES spectra for the $x=$ 0 sample by varying the number of first nearest Sr atoms. Insets (a1) and (a2) represent the enlarged view of the spectra in different energy regimes. (b) The Nb$_{\rm abs}$ $p$-LDOS corresponding to the XANES spectra presented in (a), and the inset shows the enlarged view of the highlighted region. (c) The comparison of simulated Nb $K$-edge spectra of the $x=$ 0 sample for all first neighbor $A$-site atoms as Sr and La (on the left axis) and corresponding Nb$_{\rm abs}$ $p$-LDOS (on the right axis), and the inset shows the enlarged view of the highlighted region. (d) The Nb $K$-edge spectra for the $x=$1 sample with all eight first neighbor $A$-site atoms as eight Sr, four each of La and Sr as well as eight La. The insets (d1) and (d2) represent the enlarged view of the highlighted region and the corresponding Nb$_{\rm abs}$ $p$-LDOS, respectively.}
\label{XANES_Nb_K2}
\end{figure}

Therefore, we simulate the Nb $K$-edge XANES spectra for the $x=$ 0 sample by selectively removing the neighboring atoms from the spherical cluster. We find a strong dependency of feature 3 on the first nearest Sr atoms located at 3.430~\AA $\space$ from the absorbing Nb atoms, where the intensity suppressed completely on removal of all the eight first nearest Sr atoms, as show in Fig.~\ref{XANES_Nb_K2}(a) and its inset (a1). More interestingly, the tendency of doublet turning into the singlet in the EXAFS region is also evident on the removal of these Sr atoms, as shown more clearly in the inset (a2) of Fig.~\ref{XANES_Nb_K2}(a). The corresponding Nb$_{\rm abs}$ $p$-LDOS are presented in Fig.~\ref{XANES_Nb_K2}(b), where inset show the enlarged view of the highlighted (red dashed rectangle) region corresponding to feature 3. A monotonic reduction in the Nb$_{\rm abs}$ $p$-LDOS is clearly observed with the removal of first nearest Sr atoms, causing a reduced probability of the 1$s \rightarrow p$ transitions at this energy and hence decrease in the strength/intensity of feature 3. This may result from the non-local hybridization between the Nb and Sr $p$-orbitals, where an obvious reduction in the Sr $p$-LDOS is observed around this energy (see Fig.~4 of \cite{Kumar_XAS_SI}). A non-local orbital hybridization has been widely reported in several correlated electron systems; for example, the presence of Pt and Ir $d$-LDOS at the different energies cause a shift in the pre-edge peak in the Co $K$-edge spectra of La$_2$CoPtO$_6$ and La$_2$CoIrO$_6$ due to Pt/Ir 5$d$-Co 4$p$ orbital hybridization \cite{Lee_PRB_18}. 

Further, we observe a small but consistent shift in the Nb $p$-LDOS towards the lower energy with the removal of the Sr atoms [see inset of Fig.~\ref{XANES_Nb_K2}(b)], which is opposite to the observation for feature 3 in the experimental data with $x$ (see Fig.~3 of \cite{Kumar_XAS_SI}). Therefore, in order to disentangle the effect of change in the multiple scattering paths (structure) and evolution of new atomic-like states with $x$, we present in Fig.~\ref{XANES_Nb_K2}(c), the simulated Nb $K$-edge XANES spectra by replacing all the first nearest Sr atoms by La atoms in the crystal structure of the $x=$ 0 sample ($I4/m$ symmetry). A clear reduction in the intensity of feature 3, and corresponding Nb $p$-LDOS [see inset of Fig.~\ref{XANES_Nb_K2}(c)] are evident with the La substitution at the first nearest Sr site. This clearly indicate that the strength of feature 3 depends on the atomic-like states of Sr atoms. However, no shift in the Nb $p$-LDOS is observed around feature 3 with La substitution. Further, the doublet at the higher energy recovered with the La substitution, indicating  the local structure as its origin. In order to further confirm the nature of feature 3, we now use the structural parameters of $x =$1 sample ($P2_1/n$ symmetry) and simulate the Nb $K$-edge XANES spectra in three configurations: keeping (i) eight Sr, (ii) four La and four Sr, and (iii) eight La first nearest $A$-site atoms, as shown in Fig.~\ref{XANES_Nb_K2}(d). We find a reduction in the strength of feature 3 with decrease in the Sr concentration [see inset (d1)], which further indicates its dependence on the atomic like states of the Sr atoms. On the other hand, no shift in the Nb $p$-LDOS is observed with change in the nature of first nearest $A$-site atoms [see inset (d2)], however, as discussed above, a shift of $\sim$4.5~eV as compared to the $x=$ 0 sample is clearly visible (see Fig.~5 of \cite{Kumar_XAS_SI} for the direct comparison). These results suggest that unlike amplitude, the position of feature 3 depends mainly on the arrangement of Nb--La/Sr scattering paths. 

\begin{figure} 
\includegraphics[width=3.6in]{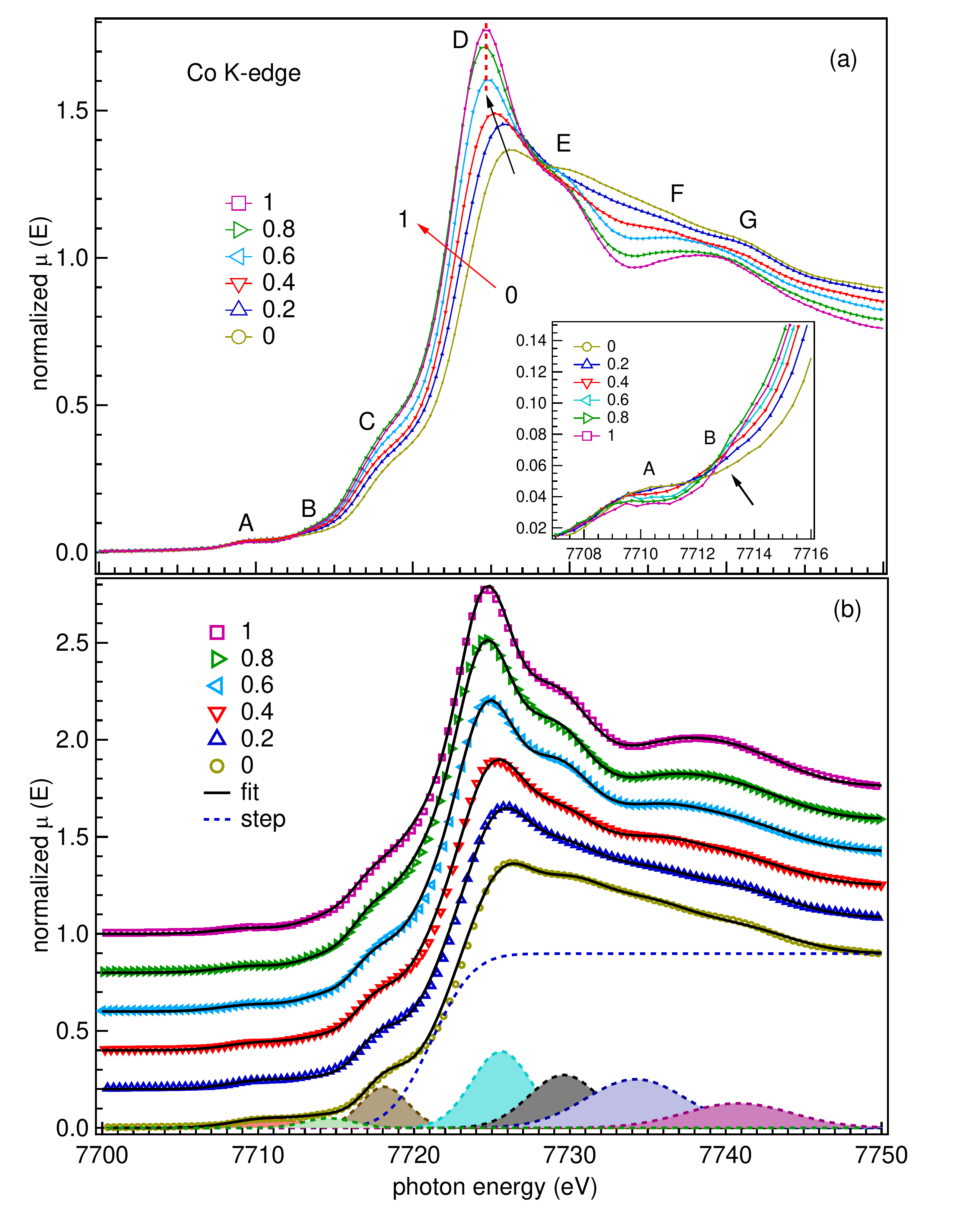}
\caption {(a) The normalized Co $K$-edge XANES spectra of Sr$_{2-x}$La$_x$CoNbO$_6$ ($x=$ 0--1) samples. The alphabets A--G represent the absorption features resulting from various electronic transitions (see text for the detail). The red arrow indicate the shift in the rising edge towards the lower energy with the La substitution. The solid black arrow and dashed red line represent the shift and invariance in the white line for x$\leqslant$0.4 and $x\geqslant$0.6 samples, respectively. Inset shows the enlarged view of the A and B (arrow) pre-edge features for the clarity. (b) The best fit of the spectra using the Gaussian peak shape and an error function as shown for the $x=$ 0 sample. Here, each spectrum is vertically shifted cumulatively by 0.2 units for the clear presentation.}
\label{XANES_Co_K}
\end{figure}

In Fig.~\ref{XANES_Co_K}(a) we show the normalized XANES spectra of Co $K$-edge, which show several absorption features (marked A to G) for Sr$_{2-x}$La$_x$CoNbO$_6$ ($x=$ 0--1) samples. The deconvolution of the different features using seven Gaussian components and an error function are presented in Fig.~\ref{XANES_Co_K}(b) and the best fit parameters are given in Table I of \cite{Kumar_XAS_SI}. The pre-edge feature A around 7710~eV can be attributed to the transitions from 1$s$ to 3$d$ states of Co, which are mixed with its 4$p$ final states and/or 2$p$ states of oxygen atoms, where the strength of this pre-edge feature increases with the extent of $p-d$ hybridization \cite{Farges_PRB_97, Wu_PRB_04, Yamamoto_XRS_08}. The presence of the strong pre-edge features in the Co $K$-edge XANES spectra suggest the higher degree of off-centre displacement of Co in the octahedra as compared to the Nb atoms (discussed above). This difference in the local coordination environment around the two B-site cations is also supported by the upcoming EXAFS analysis and useful to understand their complex electronic, magnetic, and transport properties and hence their potential use as the multiferroics \cite{Wang_AIP_13, Kumar_PRB1_20, Ishimatsu_PRB_15, Lines_Book_77}.  Further, when we increase the La concentration, a significant reduction in the pre-edge feature A is clearly observed along with a monotonic shift towards lower energy [see inset of Fig.~\ref{XANES_Co_K}(a)]. The electron doping in Sr$_{2-x}$La$_x$CoNbO$_6$ transforms Co$^{3+}$ (in $x=$ 0) in to Co$^{2+}$ (in $x=$ 1), which brings the $p$-$d$ hybridized states of Co closer to the 1$s$ states due to increase in the screening effect, resulting in the shift of this quadruple assisted transition to the lower energy. Additionally, an extra electron doped in the $d$ band of Co (3$d^6$ in $x=$ 0 and 3$d^7$ in $x=1$) results in the reduction of the density of the unoccupied $3d$ states with $x$, which reduces the transition probability from 1$s$ to partially unoccupied 3$d$-O(2$p$)/Co(4$p$) states and consequently reduction in the strength of the pre-edge feature A.

Moreover, we observe an enhancement in the intensity of feature B at around 7713~eV with the La substitution (refer table I of \cite{Kumar_XAS_SI} to see the quantitative changes), as indicated by an arrow in the inset of Fig.~\ref{XANES_Co_K}(a), also see Fig.~6 of \cite{Kumar_XAS_SI} for clarity. These two pre-edge features (A and B) have also been observed in the XANES spectra of LaCoO$_3$ and assigned as the transition from $1s$ state to the crystal field splitted unoccupied t$_{2g}$ and e$_g$ states of Co in refs.~\cite{Vanko_PRB_06, Haas_JSSC_06}. However, in the present case the Co$^{2+}$ is more likely to exist in the HS state (3$d^7$; t$_{2g}^5$e$_g^2$) as compared to Co$^{3+}$ due to weaker crystal field in the former \cite{Mabbs_book_73, Kumar_PRB1_20}, which advocate the reduction in the strength of the feature B due to decrease in the density of unoccupied e$_g$ states with the La substitution, contrary to the observation as just discussed above. This discard the transition from 1$s$ to e$_g$ states of Co as the origin of this second pre-edge feature B. In this line, based on the charge-transfer multiplet calculations, Vank\'o $et$ $al.$ assigned this second feature B as the dipolar transition from $1s$ to Co($4p$)-O(2$p$)-Co$^\prime$(3$d$) hybridize states in LaCoO$_3$, where Co$^\prime$ represent the next nearest Co ions \cite{Vanko_arxiv_12}. This hypothesis was further supported by the angular independence of the intensity of this second feature (B) in the x-ray absorption study on the epitaxial thin films of LaCoO$_3$  \cite{Sterbinsky_PRB_12}. Here, the enhancement in the strength of feature B with the La substitution indicate that latter hypothesis is clearly favored in the present case. Also, the higher transition probability from 1$s$ state to the Co($4p$)-O(2$p$)-Nb(4$d$) hybridize states as compared to Co($4p$)-O(2$p$)-Co$^\prime$(3$d$) states can be the possible reason for the enhancement in the strength of this feature B, as alternating ordering of CoO$_6$ and NbO$_6$ octahedra increases with the La substitution \cite{Kumar_PRB1_20}. 

The feature C around 7717~eV is attributed to the charge transfer from oxygen to Co ions in the final state due to strong Co 3$d$--O 2$p$ orbital hybridization, indicating the small charge transfer energy ($\Delta_{ct}$) in these compounds, i.e., $1s^1\underline c 3d^6 L^2 \rightarrow 1s^1\underline c 3d^7\underline L^1$ process, where $\underline c$ and $\underline L$ represent the core and ligand holes, respectively. For example, Chainani $et$ $al.$ found the ground state of LaCoO$_3$, an octahedrally coordinated Co$^{3+}$ perovskite, to be 38.5\% in $1s^1\underline c 3d^6 L^2$, 45.4\% in $1s^1\underline c 3d^7\underline L^1$, and 14.5\% even in $1s^1\underline c 3d^8\underline L^2$ state from their x-ray photoemission spectroscopic measurements \cite{Chainani_PRB_92}. A small but consistent reduction in the feature C is observed with the La substitution [see Table~I of \cite{Kumar_XAS_SI} and Fig.~\ref{XANES_Co_K}(b)], which indicates a decrease in the tendency of charge transfer from ligand to metal. A decrease in the valence state of Co and increase in Co--O bond distances (as evident from the EXAFS analysis, discussed later) with the La substitution can result in the enhancement of the $\Delta_{ct}$ and hence reduction in the intensity of feature C. This reduction in the orbital overlapping can also be the possible reason for decrease in the electronic conductivity of these samples with $x$ \cite {Kumar_PRB1_20}. A similar correlation between the electronic conductivity and orbital hybridization was observed from the XAS measurements of the Sr$_{2-x}$Nd$_x$Cr$_{1+x/2}$Re$_{1-x/2}$O$_6$ \cite{Blasco_PRB_07}. 

It is important to note that a monotonic shift in the rising edge (region between feature C and D) is observed towards the lower energy, as indicated by red arrow in Fig.~\ref{XANES_Co_K}(a), which indicates the conversion of Co$^{3+}$ into Co$^{2+}$ with $x$. A comparison with the Co $K$-edge XANES spectra of reference LaCoO$_3$ sample from \cite{Shukla_JPCC_21} indicate the 3+ valence state of Co in the $x=$ 0 sample, as shown in Fig.~7 of \cite{Kumar_XAS_SI}. Further, the most prominent feature D  around 1725~eV is attributed to the most probable $1s\rightarrow4p$ dipole transition. A monotonic shift in this feature to the lower energy is observed from $x=$ 0 to 0.4, as indicated by the black arrow in Fig.~\ref{XANES_Co_K}(a), whereas this feature remains almost invariant with the further La substitution for $x\geqslant$0.6 samples (vertical red dashed line). Note that the position of the white line not solely depends on the valence state but also on the local coordination environment around the absorbing atom \cite{Ishimatsu_PRB_15}. A structural transition from the tetragonal to the monoclinic phase was reported in these samples for $x\geqslant$0.6 from our previous x-ray and neutron powder diffraction measurements \cite{Kumar_PRB1_20, Kumar_NPD_22} . Thus, a cumulative effect of both valence shift and structural transformation results in the minimal shift in the feature D for the $x\geqslant$0.6 samples. Further, the strength of this feature increases 
with a monotonic reduction in the peak broadening, which indicate a decrease in the disorders in these samples with $x$. This is because the disorders result in the statistically varying energy states, which broadens the spectra, whereas $f$-sum rule states that the overall integrated intensity of the absorption spectra remains constant, irrespective of the final state \cite{Altarelli_PRB_72}, causing a reduction in the peak height \cite{Cuartero_PRB_16}. A similar variation in the nature of the white line has been reported in the Ti $K$-edge absorption spectra of various complex oxides with the varying degree of the disorders in \cite{Farges_PRB_97}, where a comparison between the crystalline, radiation damaged, and glassy CaTiSiO$_5$ clearly demonstrate the increase in the broadening and decrease in the main-edge peak height with reduction in the crystallinity \cite{Farges_PRB_97}. A variation in the white line of the Co $K$-edge XANES spectra of Sr$_{2-x}$La$_x$CoNbO$_6$ is more clearly illustrated in Fig.~8 of \cite{Kumar_XAS_SI}. In the present case, reduction in the octahedral distortion (in CoO$_6$) as well as B-site disorders with $x$ can be the possible origin for this monotonic enhancement in the intensity of the white line of Co $K$-edge XANES spectra. On the other hand, the Nb $K$-edge XANES spectra show no significant variation in the intensity of the white line up to $x=$ 0.4 [see inset (a1) of Fig.~\ref{XANES_Nb_K}(a)]. Note that the influence of the B-site disorders on the main-edge feature is expected to remain same for both Co and Nb $K$-edge spectra. This clearly demonstrates that the intensity of white line is solely  governed by the degree of the octahedral distortion in these compounds, which is also evident from the EXAFS analysis, discussed later. 

Further, we find the decrease in strength of post edge features E, F, and G with $x$. For example, the evolution of a valley between E and F features is clearly observed in Fig.~\ref{XANES_Co_K}(a). Using spin-polarized DOS calculations, Pandey $et$ $al.$ assigned the similar post edge features in the Co $K$-edge XANES spectra of LaCoO$_3$ as the transition of Co 1$s$ electrons to Co 4$p$ states hybridized with the La 6$p$ and/or O 2$p$ states \cite{Pandey_JPCM_06}. In the present case, our analysis of the FEFF simulations suggest that a decrease in the O 2$p$ states in the post edge region causes a reduction in theses features (particularly feature F) with the La substitution (see Fig.~9 of \cite{Kumar_XAS_SI}). Here it is important to re-emphasize the fact that  transformation of Co from 3+ to 2+ state with the La substitution decreases the extent of orbital overlapping between Co and oxygen atoms due to lower CF in the latter. This reduces the delocalized nature of the Co $p$ states and hence the strength of post-edge features in the Co $K$ edge XANES spectra with $x$. 

\begin{figure*} 
\includegraphics[width=7in]{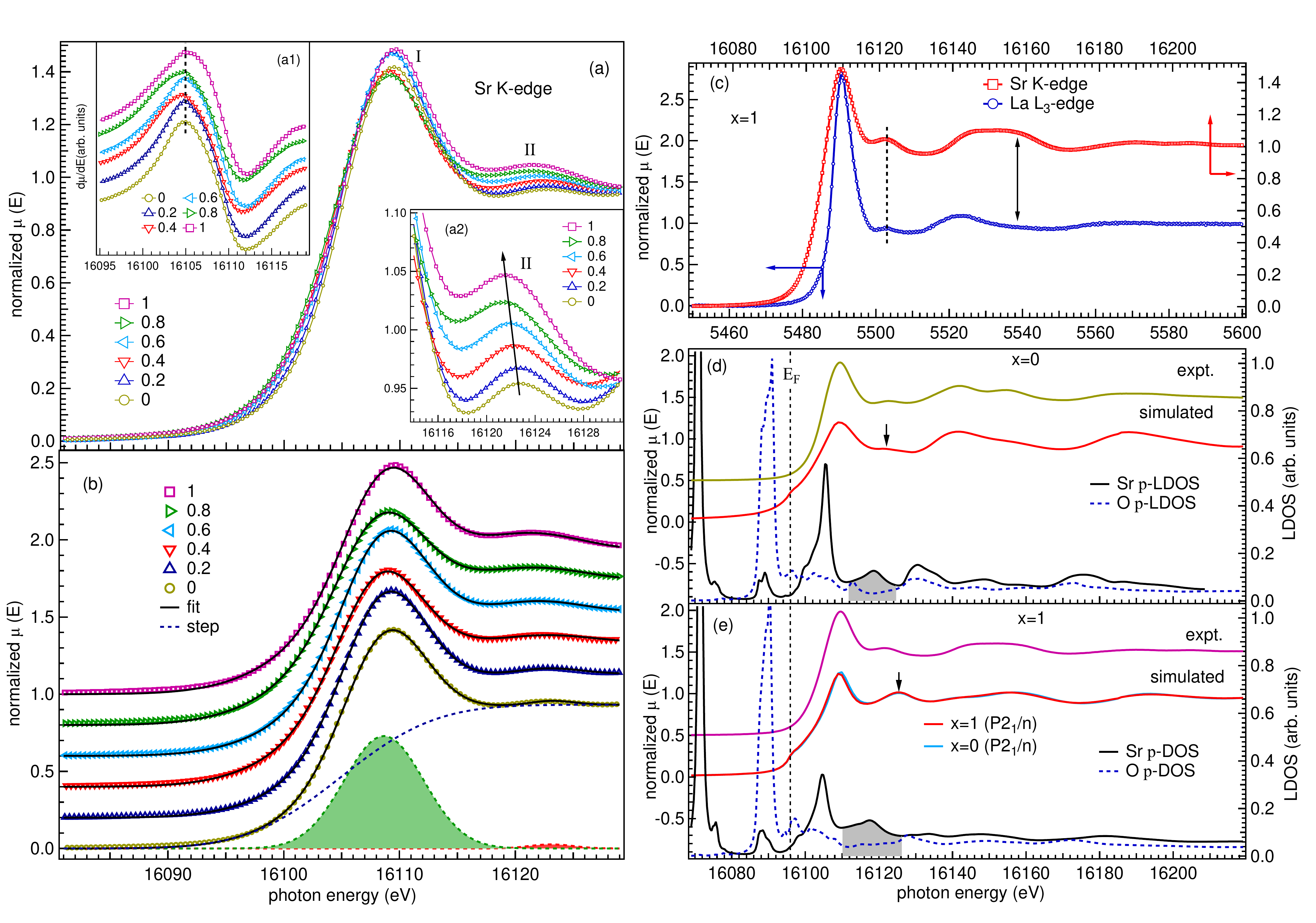}
\caption {(a) The normalized Sr $K$-edge XANES spectra of Sr$_{2-x}$La$_x$CoNbO$_6$ ($x=$ 0--1) samples. The inset (a1) shows the first derivative of the absorption coefficient, where the vertical dashed line represents the invariance in the rising edge position. These curves are vertically shifted for better presentation. The inset (a2) presents the enlarged view of feature II, where arrow shows the shift in the peak position with $x$. (b) The best fit of the Sr $K$-edge where each spectrum is vertically shifted cumulatively by 0.2 units for the clear presentation. (c) The normalized Sr $K$- (on top and right axis) and La $L_3$-edge (on bottom and left axis) spectra for the $x=$ 1 sample in 150~eV range in the vicinity of the white line. The vertical dashed line represents the presence of the absorption feature II at the same position with respect to the white line in both the spectra. The experimental and simulated Sr $K$-edge absorption spectra for the (d) $x=$ 0 and (e) $x=$ 1 samples along the calculated Sr and O $p$-LDOS. A simulated Sr $K$-edge spectrum of the $x=$ 0 sample with the structural parameters of the $x=$ 1 is compared in (e). The experimental spectra in (d, e) have been vertically shifted by 0.5 units for clarity. The shaded regions in (d) and (e) represent the unoccupied Sr $p$-LDOS corresponding to the feature II.} 
\label{XANES_Sr_K}
\end{figure*}

Now we move to the discussion of the Sr $K$-edge XANES spectra [see Fig.~\ref{XANES_Sr_K}(a)] to probe the chemical environment and local structure around the $A$-site atoms and its evolution with La substitution. There is no signature of pre-edge peak even in the first derivative of the absorption coefficient (not shown) for all the samples. The inset (a1) of Fig.~\ref{XANES_Sr_K}(a) shows the first derivative of the absorption coefficient in the rising edge region, where no significant shift (as indicated by the vertical dashed line) indicates the invariance in the valence state of Sr with $x$. Further, the most intense feature, marked as I, is attributed to the dipole allowed $1s\rightarrow 5p$ transitions, which slightly increases with $x$. Interestingly, the intensity of post edge feature II increases significantly with $x$, as shown more clearly in the inset (a2) of Fig.~\ref{XANES_Sr_K}(a). Also, a shift in feature II towards lower energy is clearly observed with $x$, as shown by the arrow in inset (a2). In Fig.~\ref{XANES_Sr_K}(b), we show the the best fit XANES spectra for all the samples and the best fit parameters are given in Table I of \cite{Kumar_XAS_SI}. A similar post edge feature in the La L$_3$ and Sr $K$ edge XANES spectra of La$_{2-x}$Sr$_x$CuO$_4$ samples is attributed to the removal of apical oxygen around the absorbing atoms with Sr substitution \cite{Tan_PRL_90}. However, this hypothesis was later contradicted by the theoretical calculations by Wu $et$ $al.$ \cite{Wu_PRBR_92}. In order to understand the origin of this feature, 
La L$_3$-edge XANES spectrum has been recorded for the $x=$ 1 sample and compared with Sr $K$-edge in Fig.~\ref{XANES_Sr_K}(c). The presence of similar absorption feature above the white line in case of La L$_3$-edge XANES spectra 
[see vertical dashed line in Fig.~\ref{XANES_Sr_K}(c)] discard the atomic-like states of Sr as the origin of this feature. This suggest that the multiple scattering from the neighboring atoms give rise to the feature II in these samples. Further, a comparison of the La L$_3$-edges spectra of the $x=$ 1 sample with the reference La$_2$O$_3$ sample indicate the presence of La in the trivalent state (see Fig.~10 of \cite{Kumar_XAS_SI}).

In order to further understand the origin of this feature II, we simulate the Sr $K$-edge XANES spectra for the $x=$ 0 and 1 samples for a cluster of 45 atoms and present in Figs.~\ref{XANES_Sr_K}(d, e), respectively. A close synchronization with the experimental spectra is observed after shifting the simulated spectra by -25.5~eV for both the samples. A clear enhancement in the post edge feature II is visible in the simulated spectrum of the $x=$ 1 sample as compared to the $x=$ 0 sample, marked by the arrows in Figs.~\ref{XANES_Sr_K}(d, e). Further, we simulate the Sr $K$-edge XANES spectrum of the $x=$ 0 sample using the structural parameters of the $x=$ 1 sample ($P2_1/n$) and compare in Fig.~\ref{XANES_Sr_K}(e). We find no significant change in the intensity of the feature II, which further confirms its structural origin. Our simulations performed by the selective removal of the neighboring atoms indicate that the feature II in the Sr $K$-edge XANES spectra results from the multiple scattering from the 12 nearest neighbor oxygen atoms, where any change in the Sr--O scattering paths with the La substitution alter the oxygen $p$-LDOS at this energy, as shown in Figs.~\ref{XANES_Sr_K}(d, e) and hence intensity and position of the feature II. However, the observed pre-edge shoulder and insignificant shift in the position of feature II in the simulated spectra are contrary to the experimental observation for both the samples. Here, it is important to note that the input structural parameters for the simulation were used from the XRD/NPD data, which gives an average crystal structure of the samples. The details of the local structure around the absorbing atom are missing, because unlike element specific XAS technique, diffraction techniques are site specific. For example, in case of the $x=$ 1 sample, an average distortion in the SrO$_{12}$ and LaO$_{12}$ polyhedra can be probed by the diffraction measurements \cite{Bos_PRB_04}, the spectral features of Sr $K$-edge on the other hand depends solely on change in the local coordination around the Sr atoms. These can be possible reasons for the observed discrepancy between the experimental and simulated spectra in Figs.~\ref{XANES_Sr_K}(d, e). 

\begin{figure*} 
\includegraphics[width=7.03in]{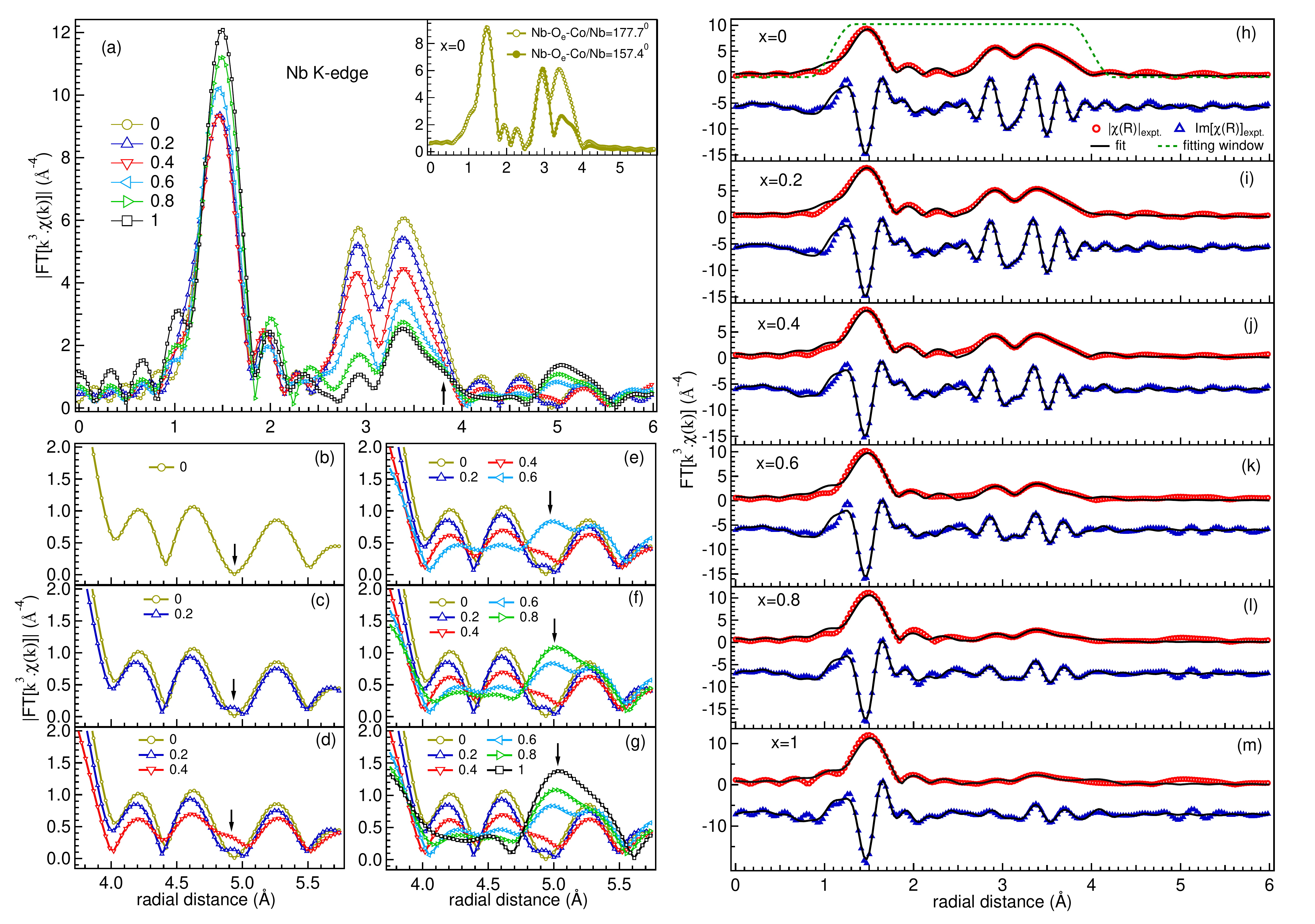}
\caption {(a) The magnitude of the Fourier transformed $k^3 \chi(k)$ spectra for Sr$_{2-x}$La$_x$CoNbO$_6$ ($x=$ 0--1) in 0--6 \AA $\space$ spectral range, derived from the EXAFS region of the Nb K-edge. The inset shows the comparison of simulated spectra of the $x=$ 0 sample for two different Nb--O$_e$--Co/Nb angles (Nb--O$_a$--Co/Nb=180$^0$). (b)--(g) The enlarged view of (a) between 3.75 to 5.75 \AA $\space$ with cumulative increment in $x$ to clearly show the evolution of the new scattering path resulting from the Co/Nb atoms, located diagonally to the absorber, as indicated by the arrows (refer text for more details). (h--m) The fitting of the $|\chi(R)|$ and corresponding Im[$\chi(R)$] spectra in the window shown by the dotted green line, each Im[$\chi(R)$] spectrum is shifted downward for clarity in presentation.}
\label{DP_Nb_K}
\end{figure*}

\begin{table*}
	
		\label{Table_fit}
	\caption{The EXAFS fitting parameters for Sr$_{2-x}$La$_x$CoNbO$_6$ samples extracted from the Nb, Co, and Sr $K$-edge spectra.}

\begin{tabular}{p{4cm}p{2cm}p{2cm}p{2cm}p{2cm}p{2cm}p{2cm}p{1.9cm}p{2cm}p{2cm}}
\hline
\hline

$x$ &0&0.2&0.4&0.6& 0.8&1\\

          \hline

&&&Nb K-edge&&&\\
\hline

	       R$_{\rm Nb-O_e}$ (\AA)  & 2.024(2) & 2.025(3) & 2.038(1) & 2.028(4) & 2.020(1) & 2.023(2)\\
               $\sigma^2_{\rm Nb-O_e}$ (x10$^{-3}$\AA$^2$) & 4.1(3) & 4.5 (1) & 5.7(2) & 4.7(1) & 4.1(1) & 4.1(1)\\

  R$_{\rm Nb-O_a}$ (\AA)  & 1.914(2) & 1.915(3) & 1.924(4) & 1.925(1) & 1.928(1) & 1.939(2)\\

               $\sigma^2_{\rm Nb-O_a}$ (x10$^{-3}$\AA$^2$) & 0.2(2) & 0.3(2) & 0.2(2) & 0.2(4) & 0.3(2) & 0.2(3)\\
          
$\Delta_{\rm Nb}$x10$^{-4}$  & 6.8(2) & 6.8(3) & 7.2(4) & 5.9(4) & 4.8(1) & 3.9(2)\\

  R$_{\rm Nb-Sr}$ (\AA)  & 3.431(2) & 3.436(2) & 3.435(4) & 3.451(1) & 3.436(1) & 3.498(4)\\

              $\sigma^2_{\rm Nb-Sr/La}$ (x10$^{-3}$\AA$^2$) & 7.8(1) & 7.5(3) & 8.5(4) & 9.6(1) & 7.9(1) & 10.3(2)\\

  R$_{\rm Nb-La}$ (\AA)  & - & 3.439(2) & 3.480(6) & 3.460(1) & 3.426(3) & 3.481(4)\\

  R$_{\rm Nb-O-Nb-Nb}$ (\AA)  & 3.771(2) & 3.782(3) & 3.781(1) & 3.806(4) & 3.722(3) & 3.822(6)\\

$\sigma^2_{\rm Nb-O-Nb-Nb}$ (x10$^{-3}$\AA$^2$) & 5.4(1) & 4.4(3) & 6.7(4) & 4.6(2) & 6.3(5) & 4.5(5)\\

N$^{Nb}$ & 3.0(1) & 2.9(0) & 2.7(1) & 1.2(1) & 0.7(0) & 0.3(1)\\

$S$ & 0.00(3) & 0.03(1) & 0.10(3) & 0.60(3) & 0.77(1) & 0.90(3)\\

 R factor & 0.013 & 0.013 & 0.017 & 0.027 & 0.019 & 0.019\\

\hline

&&&Co K-edge&&&\\
\hline
  R$_{\rm Co-O_e}$ (\AA)  & 1.925(3) & 1.932(4) & 1.935(1) & 1.968(2) & 2.020(6) & 2.023(1)\\

               $\sigma^2_{\rm Co-O_e}$ (x10$^{-3}$\AA$^2$) & 8.9(2) & 8.3(4) & 7.6(2) & 6.3(1) & 4.5(4) & 4.4(1)\\

  R$_{\rm Co-O_a}$ (\AA)  & 2.059(2) & 2.059(3) & 2.058(2) & 2.085(1) & 2.134(5) & 2.126(1)\\

 $\sigma^2_{\rm Co-O_a}$ (x10$^{-3}$\AA$^2$) & 5.7(2) & 4.6(4) & 2.6(2) & 1.8(1) & 1.5(2) & 2.1(1)\\

$\Delta_{\rm Co}$x10$^{-4}$  &10.3(3) & 9.2(4) & 8.6(2) & 7.5(2) & 6.8(6) & 5.6 (1)\\

  R$_{\rm Co-Sr}$ (\AA)  & 3.655(2) & 3.676(4) & 3.623(1) & 3.551(1) & 3.541(3) & 3.517(1)\\

               $\sigma^2_{\rm Co-Sr/La}$ (x10$^{-3}$\AA$^2$) & 11.6(2) & 10.1(5) & 18.2(3) & 18.0(6) & 18.6(3) & 17.5(1)\\

 R$_{\rm Co-La}$ (\AA)  & -& 3.640(4) & 3.693(1) & 3.651(1) & 3.633(5) & 3.650(1)\\

R$_{\rm Co-O-Co-Co}$ (\AA)  & 3.899(2) & 3.889(3) & 3.835(1) & 3.781(4) & 3.776(7) & 3.937(2)\\

$\sigma^2_{\rm Co-O-Co-Co}$ (x10$^{-3}$\AA$^2$) & 5.4(1) & 7.4(3) & 8.3(1) & 4.9(2) & 4.2(5) & 5.0(5)\\

 R factor & 0.007 & 0.024 & 0.017 & 0.013 & 0.029 & 0.015\\

            \hline

&&&Sr K-edge&&&\\
\hline

         R$_{\rm Sr-O_1}$ (\AA) & 2.560(2) & 2.543(2) & 2.534(1) & 2.535(3) & 2.522(4) & 2.519(1)\\
         R$_{\rm Sr-O_2}$ (\AA) & 2.668(2) & 2.670(1) & 2.671(2) & 2.618(2) & 2.639(3) & 2.563(2)\\
         R$_{\rm Sr-O_3}$ (\AA) & 2.857(3) & 2.848(1) & 2.870(2) & 2.844(2) & 2.837(4) & 2.803(1)\\
        
$\sigma^2_{\rm Sr-O_1/O_2/O_3}$ (x10$^{-3}$\AA$^2$) & 9.2(3) & 6.9(2) & 8.7(3) & 11.5(1) & 10.3(4) & 13.6(3)\\

       R$_{\rm Sr-Co/Nb}$  (\AA) & 3.315(2) & 3.327(2) & 3.347(1) & 3.415(2) & 3.451(5) & 3.453(2)\\

$\sigma^2_{\rm Sr-Sr/La}$ (x10$^{-3}$\AA$^2$) & 5.8 (1) & 8.8(2) & 12.1(2) & 12.7(3) & 13.0(4) & 11.1(1)\\

        R$_{\rm Sr-Sr/La}$ (\AA)  & 3.473(2) & 3.482(1) & 3.483(2) & 3.486(1) & 3.463(5) & 4.500(2)\\
      $\sigma^2_{\rm Sr-Sr/La}$ (x10$^{-3}$\AA$^2$) & 8.8(1) & 8.8 (2) & 9.1(2) & 11.4(3) & 11.8(4) & 16.2(1)\\

                  R factor & 0.012 & 0.014 & 0.008 & 0.011 & 0.019 & 0.010\\
\hline
\hline
\end{tabular}
\end{table*}

\subsection{\noindent ~Extended x-ray absorption fine structures:}

Having discussed the electronic structure, we now focus on the EXAFS part of XAS spectra to understand the local coordination environment around different elements in these samples. In Fig.~\ref{DP_Nb_K}(a), we show the amplitude of the Fourier transformed $k^3 \chi(k)$ of high-resolution EXAFS spectra of Nb $K$-edge for all the samples and the normalized raw data are presented in Fig.~3 of \cite{Kumar_XAS_SI}. The background-subtracted EXAFS spectra are fairly smooth up to $k\approx$14 \AA$^{-1}$ (see Fig.~11 of \cite{Kumar_XAS_SI} for the $x=$ 0 sample), allowing us to probe even the small spectral features from the higher coordination shells with the reasonable accuracy. In Fig.~\ref{DP_Nb_K}(a), the first intense maxima around 1.5 \AA $\space$ (phase uncorrected) is attributed to the Nb--O scattering paths in the NbO$_6$ octahedra. It is interesting to note that the intensity of this maxima remains almost constant for $x\leqslant$ 0.4 samples and then it increases monotonically with further increase in the La substitution, analogous to the variation in intensity of the white line observed in the Nb $K$-edge XANES spectra [see inset (a1) of Fig.~\ref{XANES_Nb_K}(a)]. Here, it is important to note that in the tetragonal ($I4/m$) symmetry, the four equatorial (O$_e$; in $ab$ plane) and two axial (O$_a$; along $c$-axis) oxygen positions lead to the two sets of different Nb--O bond distances \cite{Hudson_PRB_96}. Invariance in the peak intensity corresponding to Nb--O scattering path indicates almost no change in the degree of NbO$_6$ octahedral distortion for the $x\leqslant$0.4 samples. This is also evident from the local distortion parameter, ($\Delta_d$) around the Nb atom (denoted by $\Delta_{\rm Nb}$), see Table I, quantified using below equation \cite{Zhu_PRB_20}
\begin{equation}
\Delta_d = \frac{1}{n}\sum_{i=1}^n\left(\frac{d_n-\langle d\rangle}{\langle d\rangle}\right)^2, 
\label{distortion}
\end{equation}
where $n$ is the number of atoms coordinated to that site, $d_n$ is the bond length between that site and $n$th coordinated atom, and $\langle d\rangle$ is the average bond length of all the coordinations to that site. In the monoclinic ($P2_1/n$) symmetry for $x\geqslant$0.6 samples, four Nb--O$_e$ bonds further split in two sets of different bond distances. However, in order to avoid over-parameterization, we assume the identical Nb--O$_e$ bond distances in the EXAFS fitting, as shown in Figs.~\ref{DP_Nb_K}(h--m) for the $x=$ 0--1 samples, respectively. Here, an increase in the intensity of the Nb--O scattering path for $x\geqslant$0.6 samples [see Fig.~\ref{DP_Nb_K}(a)] indicate the statistically narrow distribution of the Nb--O path lengths, i.e, decrease in the octahedral distortion (in NbO$_6$) for these samples (see $\Delta_{\rm Nb}$ values in Table I). This is found to be interesting as the NbO$_6$ octahedra accommodates the chemical pressure exerted by the substitution of smaller La$^{3+}$ ions at the Sr$^{2+}$ site \cite{Shannon_AC_76} up to the $x=$ 0.4 sample. However, with further increase in $x$, the NbO$_6$ octahedra rotates around $x$ and $y$-axis with the same magnitude, leading to the monoclinic distortion having b$^-$b$^-$c$^+$ tilt \cite{Kumar_NPD_22, Bos_PRB_04, Kumar_PRB1_20}, which results in the abrupt decrease in the distortion of NbO$_6$ octahedra for the $x\geqslant0.6$ samples. This is also manifested by the sudden enhancement in the B-site ordering (refer Table I and discussion below) and hence antiferromagnetic interactions in $x\geqslant0.6$ samples through Co--O--Nb--O--Co 180$^0$ superexchange path \cite {Kumar_PRB1_20}. 

\begin{figure*} 
\includegraphics[width=7.2in]{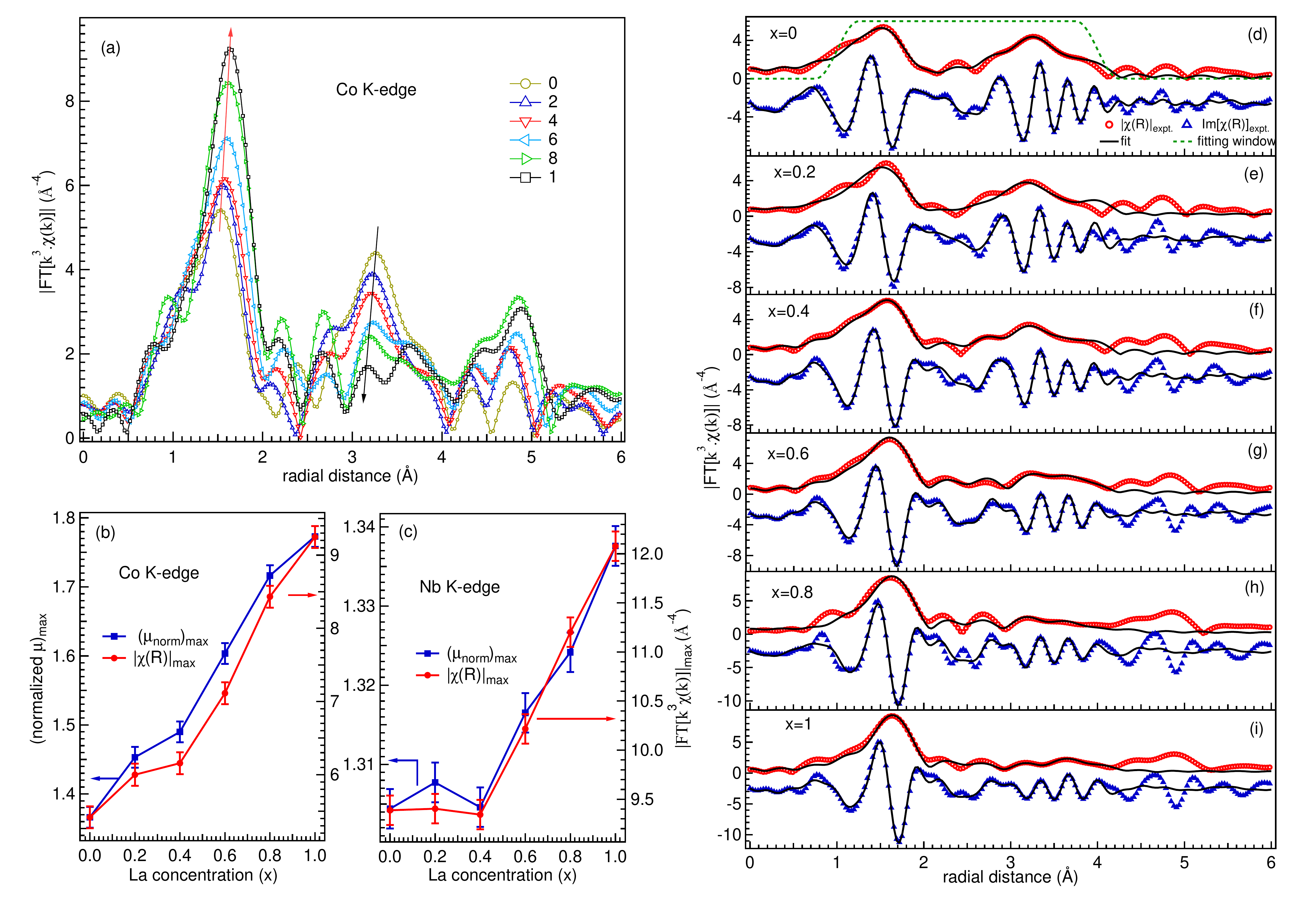}
\caption {(a) The magnitude of the Fourier transformed $k^3\chi(k)$ spectra for Sr$_{2-x}$La$_x$CoNbO$_6$ ($x=$ 0--1) in 0--6 \AA $\space$ spectral range, derived from the EXAFS spectra of Co $K$-edge. (b) The evolution of the intensity of the white line (left axis) and amplitude of the first coordination shell (right axis) with the La substitution for the Co $K$-edge and (c) Nb $K$-edge EXAFS spectra. (d--i) The fitting of the $|\chi(R)|$ and corresponding Im[$\chi(R)$] spectra in the window shown by the dotted green line, where the Im[$\chi(R)$] spectrum in each panel is shifted downward for the clarity in the presentation.} 
\label{DP_Co_K}
\end{figure*}

The second and third maxima at around 2.9 and 3.4 \AA $\space$ are attributed to the Nb--Sr/La and Nb--Co/Nb scattering paths, respectively. Interestingly, the intensity of both the scattering paths suppresses with $x$, where Nb--Sr/La path decreases more rapidly as compared to Nb--Co/Nb path [see Fig.~\ref{DP_Nb_K}(a)]. The La substitution at Sr site results in the evolution of new Nb--La scattering paths at a different radial distance as compared to Nb--Sr paths. Also, a structural distortion [from tetragonal ($I4/m$) to monoclinic ($P2_1/n$) phase] resulting from the La substitution causes splitting in the eight equal Nb--Sr scattering paths, and consequently a reduction in the strength of the corresponding feature in $|$FT[$k^3\chi$(k)]$|$ plot. However, the EXAFS is unable to resolve these spatially close ($\Delta r<$0.02) Nb--Sr/La scattering paths and such a close statistical distribution in the path lengths is incorporated in the enhancement of Debye-Waller (DW) factor ($\sigma^2$) corresponding to that path. Here, we assume the same DW factor for Nb--Sr and Nb--La scattering paths to fit the Nb $K$-edge EXAFS spectra. A small enhancement in the Nb--Sr/La scattering path length is also observed with $x$ [see Fig.~\ref{DP_Nb_K}(a)]. Here, a competition between decrease in the ionic radii of $A$-site cations (r$_{\rm La^{3+}}$=1.36~\AA $\space$ and r$_{\rm Sr^{2+}}$=1.44~\AA; 12 coordinated \cite{Shannon_AC_76}) and increase in the Coulombic repulsion between Nb$^{5+}$--Sr$^{2+}$/La$^{3+}$ ions with $x$ govern this scattering path length and the dominance of the latter effect results in the observed enhancement in this scattering path length with $x$. Further, the Nb--Co/Nb scattering paths are mediated by the oxygen atoms and hence forward scattering paths Nb--O--Nb/Co--Nb are more favorable as compared to the single scattering (SS) Nb--Co/Nb paths. The x-ray diffraction patterns clearly evident that structural distortion due to La substitution causes significant deviation in the Nb--O--Co/Nb bond angle (from $\approx$179$^0$ in $x=$ 0 to $\approx$158$^0$ in the $x=$ 1 sample \cite{Kumar_PRB1_20}), which transform Nb--O--Nb/Co--Nb forward scattering path into an obtuse triangle with the lower scattering amplitude. This results in the reduction in both the MS paths through Nb atoms, causing the observed monotonic reduction in the corresponding peaks in $|$FT[$k^3\chi (k)]|$ plot [see Fig.~\ref{DP_Nb_K}(a)]. In order to confirm this, we simulate the $|$FT[$k^3\chi(k)]|$ spectra for the $x=$ 0 sample by varying the Nb--O$_e$--Co/Nb angle from 177.7$^{\rm o}$ to 157.4$^{\rm o}$, keeping all the other parameters constant, as shown in the inset of Fig.~\ref{DP_Nb_K}(a). A significant reduction in the spectral weight around 3.5~\AA~ clearly indicate that the oxygen mediated MS path is the origin of this feature. A similar change in the pseudo-radial distribution function is reported in Sr$_{2-x}$Ca$_x$YIrO$_6$ samples with change in the Y--O--Ir angle \cite{Marco_PRB_20}. Also, it is important to note that in a completely B-site disordered structure, each Nb atom is statistically coordinated with 3 Co and 3 Nb first nearest B-site neighbors, whereas in case of fully ordered structure, each Nb atom is coordinated with 6 Co atoms \cite{Das_PRB_20, Bandyopadhyay_PRB_17}. Here, the coordination number of the first nearest Nb atoms ($N_{\rm Nb}$) and the resulting order parameter, $S$=(3--$N_{\rm Nb}$)/3 from the best fit of the EXAFS spectra are listed in Table I, where $S=$ 0 and 1 represent the completely disordered and ordered structures, respectively. A clear enhancement in the B-site ordering with the La substitution is evident from Table I. We use the same zero-energy shift parameter ($\Delta$E$_0$) and many-body amplitude reduction factor (S$_0^2$=0.8) for all the scattering paths to fit the Nb $K$-edge EXAFS spectra of Sr$_{2-x}$La$_x$CoNbO$_6$ samples. Moreover, we consider a single scattering path Nb--O2 (where O2 indicate the second nearest oxygen atoms; see Fig.~12(a) of \cite{Kumar_XAS_SI}) at around 3.8 \AA $\space$ causing an asymmetry in the peak at the higher radial distance, as indicated by the arrow in Fig.~\ref{DP_Nb_K}(a). This asymmetry appears to become more prominent as the intensity of the main peak at 3.4 \AA~reduces with increase in the La substitution. 

More importantly, a close inspection in the range of $\sim$4--5.6 \AA $\space$ clearly indicate the evolution of new spectral features, as shown in Figs.~\ref{DP_Nb_K}(b--g) cumulatively with $x$. Here, the feature around 4.6 \AA~is resulting from the Nb--La/Sr-O2--Nb double scattering path. As discussed above, the splitting in Nb--Sr/La scattering paths with the La substitution provide the statistically wide range of path lengths for their corresponding MS paths, causing the observed consistent reduction in the intensity of this peak in the $|$FT[$k^3 \chi(k)]|$ plot. Further, a feature around 5.3 \AA $\space$ is resulting from the Nb--O3 single scattering path, as shown in Fig.~12(b) of \cite{Kumar_XAS_SI}. Another interesting observation is the evolution of a new feature around 5 \AA $\space$ with the La substitution, as indicated by arrows in Figs.~\ref{DP_Nb_K}(b--g). This new feature is resulting from the single scattering from second nearest B-site atoms (Nb--Co2/Nb2), i.e., along $<110>$ directions of pseudocubic unit cell, as shown in Figs.~12(c, d) of \cite{Kumar_XAS_SI}. The completely random occupancy of Co and Nb atoms at two Wyckoff positions in case of the $x=$ 0 sample and its minimal deviation from the cubic symmetry results in the absence of this feature in \ref{DP_Nb_K}(b), as Nb--Co2/Nb2 SS path is mediated by the central Sr atoms in the perfectly symmetric environment. Further, a reduction in the crystal symmetry with the La substitution, as also speculated from the deviation in the tolerance factor ($\tau$) from unity \cite{Kumar_PRB1_20}, results in the evolution of this feature with $x$, as shown in Figs.~\ref{DP_Nb_K}(b--f). This is consistent with the evolution of new Raman active modes in these samples with the La substitution \cite{Kumar_PRB1_20}. Further, an abrupt increase in the strength of this feature for the $x\geqslant$0.6 samples [(see Figs.~\ref{DP_Nb_K}(b--g)] is consistent with the structural transition from tetragonal ($I4/m$) to the monoclinic ($P2_1/n$) structure accompanied by the sharp enhancement in the B-site ordering (see Table I) at this critical doping ($x=$ 0.6), which is correlated with the degree of octahedral distortion in the NbO$_6$ unit, as discussed above. Though the consistent variation in these features with the La substitution up to the radial distance of around 6 \AA $\space$ (phase corrected) clearly indicate the high data quality, we fitted the EXAFS spectra upto 4 \AA$\space$ to improve the reliability in the extracted parameters. It is important to note that despite the overall enhancement in the structural distortion, the distortion in the NbO$_6$ octahedra decreases  with the La substitution. 

\begin{table*}
		\label{Table_fit}
		\caption{The charge transfer ($\Delta$e) and charge counts in the $s$, $p$, and $d$ states of Nb$_{\rm abs}$ and O atoms extracted from the simulation of the Nb $K$-edge XANES spectra of the $x=$ 0 sample by varying Nb--O bond distances.}

\begin{tabular}{p{2cm}p{1.5cm}p{1.5cm}p{1.5cm}p{1.5cm}p{1.5cm}p{1.5cm}p{1.5cm}p{1.5cm}p{1.5cm}}
\hline
\hline

Nb--O$_a({\rm \AA})\rightarrow$  & & 1.990 & 1.980 &	1.970&	1.960&	1.950&	1.940&	1.930&	1.920\\
Nb--O$_e({\rm \AA})\rightarrow$  & &1.990 &	2.000 &	2.010&	2.020&	2.030&	2.040&	2.050&	2.060\\
\hline

 Nb$_{\rm abs}$&   $s$  & 0.392&	0.392&	0.391&	0.390&	0.389&	0.387&	0.386&	0.385\\
Nb$_{\rm abs}$ & $p$ &6.603&	6.602&	6.601&	6.600&	6.598&	6.598&	6.597&	6.5967\\
Nb$_{\rm abs}$ & $d$& 4.143&	4.151&	4.164&	4.175&	4.184&	4.194&	4.204&	4.213\\
Nb$_{\rm abs}$ & $\Delta$e  & 0.861&	0.855&	0.844&	0.836&	0.828&	0.821&	0.813&	0.806\\
O &   $s$  & 1.835&	1.836&	1.835&	1.835&	1.835&	1.834&	1.834&	1.834\\
O & $p$ &4.498&	4.496&	4.492&	4.488&	4.485&	4.481&	4.477&	4.473\\
O & $d$ & 0.074&	0.075&	0.076&	0.077&	0.078&	0.079&	0.080&	0.081\\
O &  $\Delta$e  &-0.408&	-0.407&	-0.403&	-0.400&	-0.397&	-0.394&	-0.391&	-0.387\\
   
\hline
\hline
\end{tabular}
\end{table*}

Now we discuss the effect of La substitution on the local coordinate environment around the Co atoms. In Fig.~\ref{DP_Co_K}(a), we show the amplitude of the Fourier transformed $k^3 \chi(k)$ of Co $K$-edge EXAFS spectra for all the samples. The normalized raw data are plotted in Fig.~13 of \cite{Kumar_XAS_SI} along with the other details. The most prominent feature around 1.5~\AA $\space$ results from the Co--O scattering paths. The intensity of this feature increases monotonically with the La substitution, unlike Nb--O scattering path in the Nb $K$-edge EXAFS spectra, which remains constant up to the $x=$ 0.4 [see Fig.~\ref{DP_Nb_K}(a)]. This suggest a different evolution of local coordination environment around the Co and Nb atoms with the La substitution. Here a higher degree of octahedral distortion results in the low intense wider peak in the $|\chi(R)|$ distribution function as compared to the Nb $K$-edge spectra. Importantly, the increase in the intensity of this feature at 1.5~\AA $\space$ is consistent with the variation in the strength of white line of Co $K$-edge XANES spectra [see Fig.~\ref{XANES_Co_K}(a)], analogous to the correlation observed in the Nb $K$-edge. The fitted Co $K$-edge EXAFS spectra are presented in Figs.~\ref{DP_Co_K}(d--i) and the best fit parameters are provided in Table I. We observe the $z$-out distortion in the CoO$_6$ octahedra, which decreases with $x$, inspite of the reduction in the crystal symmetry (see $\Delta_{\rm Co}$ values in Table I). This indicate the presence of electronic instability assisted Jahn-Teller (JT) distortion in these samples, as the intermediate spin (IS) state of Co$^{3+}$ ($t_{2g}^5$e$_g^1$) is JT active \cite{Yamaguchi_PRB_97, Shukla_JPCC_21}. This $z$-out distortion in the CoO$_6$ octahedra is expected to cause the observed $z$-in distortion in the neighboring NbO$_6$ octahedra (see Table I). Further, a decrease in the DW factor of the Co--O scattering paths is observed with $x$, suggesting the reduction in both static as well as dynamic distortion around the Co atoms. An enhancement in the concentration of Co$^{2+}$ ions with the La substitution results in the reduction of this $z$-out JT distortion in the CoO$_6$ octahedra, as Co$^{2+}$ is more likely to exist in the JT inactive HS state as compared to Co$^{3+}$ \cite{Mabbs_book_73, Kumar_PRB1_20}. However, this simplified hypothesis based on the pure ionic picture may partially breakdown due to the covalent character of Co--O bond, resulting from the higher crystal field in these samples. For example, an increase in the  occupancy of 3$d$ states (more than 6 electrons for Co$^{3+}$) and the partial occupancy of e$_g$ states has been reported even for the LS configuration of LaCoO$_3$ \cite{Pandey_PRB_08, Hsu_PRB_09}. Note that the increase in ionic radius of Co (0.545\AA $\space$ for LS and 0.610~\AA $\space$ for HS Co$^{3+}$, and 0.650~\AA $\space$ for LS and 0.745~\AA $\space$ for HS Co$^{2+}$; 6 coordinated \cite{Shannon_AC_76}) and decrease in the Coulombic attraction between oxygen and Co atoms results in the shift of Co--O scattering part to the higher radial distance with the La substitution, as shown by the red arrow in Fig.~\ref{DP_Co_K}(a). Further, a feature around 3.2~\AA $\space$ is attributed to the Co--La/Sr scattering path. Here, we find that the strength of this feature as well as the scattering path length reduces with $x$, as shown by the black arrow in Fig.~\ref{DP_Co_K}(a). However, note that the  strength decreases, but the scattering path length increases with $x$ in case of Nb--La/Sr, as observed and discussed above. Here, a cumulative effect of increase in the ionic radii of Co (Co$^{3+}$ $\rightarrow$Co$^{2+}$), decrease in that of $A$-site cations, and reduction in the Coulombic repulsion between Co and $A$-site cations (Co$^{3+}$--Sr$^{2+}$ for the $x=$ 0, and $<$Co$^{2+}$--Sr$^{2+}$ and Co$^{2+}$--La$^{3+}$$>$ for the $x=$ 1 sample) govern the evolution of Co--La/Sr scattering path length with the La substitution ($x$). The dominance of the latter two effects result in the observed reduction in this scattering path length with $x$, as shown in Fig.~\ref{DP_Co_K}(a). Another peak at the higher radial distance of around 3.7~\AA $\space$ is resulting from the nearest neighbor Co--Nb/Co scattering paths. The fitting of the Co $K$-edge EXAFS spectra is performed in 1--4~\AA $\space$ range, with the same procedure as described above for the Nb $K$-edge EXAFS analysis. Here, we use the same degree of the B-site ordering (i.e, fraction of the first neighbor Co atoms) as extracted from the Nb $K$-edge EXAFS spectra in order to improve the reliability in the other fitting parameters. 

\begin{figure}
\includegraphics[width=3.5in]{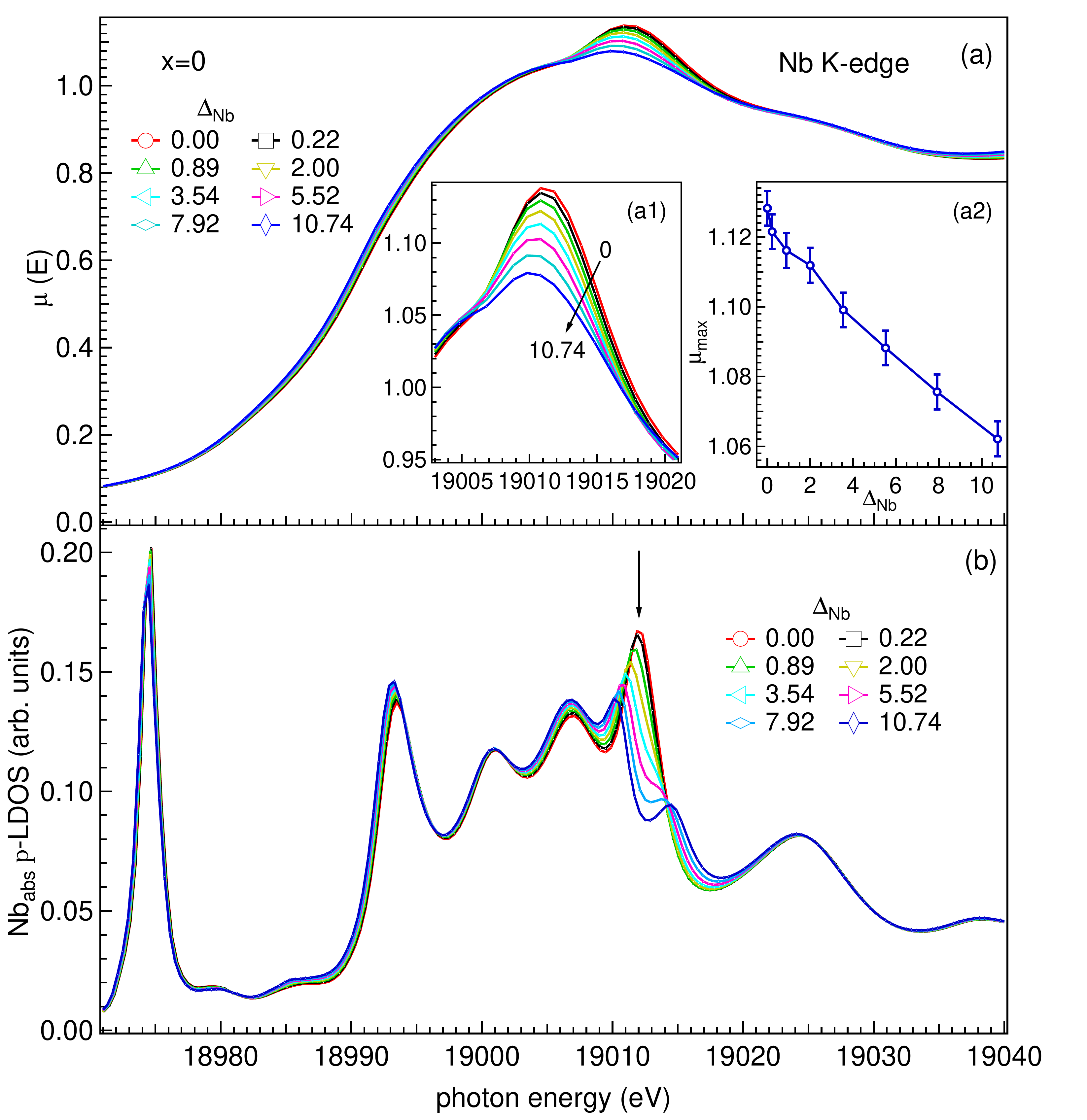}
\caption {(a) The Nb $K$-edge XANES spectra simulated for the $x=$ 0 sample by varying the degree of distortion in NbO$_6$ octahedra. The insets (a1) and (a2) represent the enlarged view of the white line and variation of the white line intensity with the distortion parameter, $\Delta_{\rm Nb}$, defined in equation \ref{distortion}, respectively. The error bars in the white line intensity in inset (a2) are accounting for the finite step size of the simulated spectra. (b) The corresponding Nb$_{\rm abs}$ $p$-LDOS simulated at different values of $\Delta_{\rm Nb}$ for the $x=$ 0 sample.} 
\label{XANES_Nb_K3}
\end{figure}

\begin{figure*} 
\includegraphics[width=7.2in]{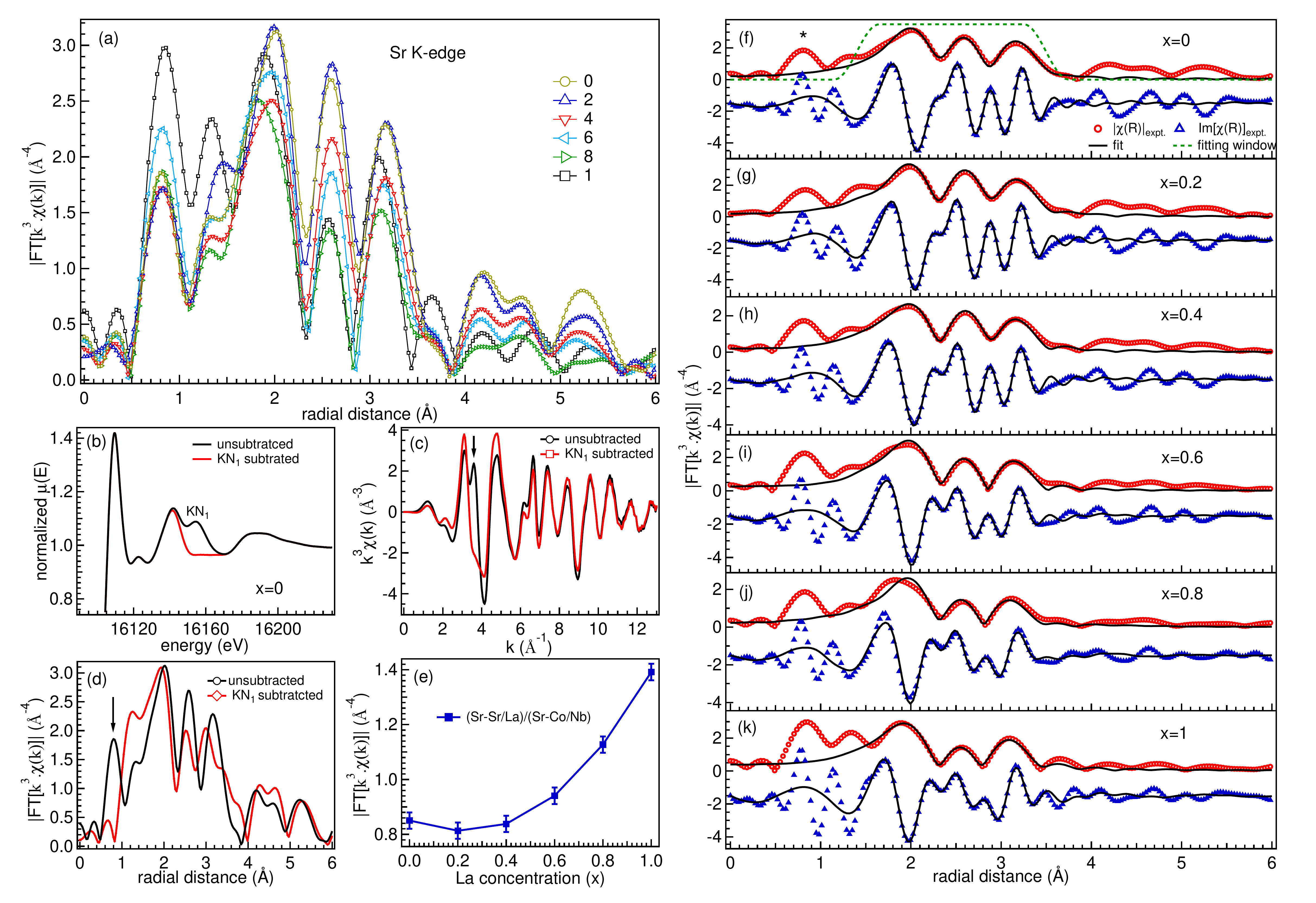}
\caption {(a) The magnitude of the Fourier transformed $k^3\chi(k)$ spectra of Sr K-edge for Sr$_{2-x}$La$_x$CoNbO$_6$ ($x=$ 0--1) samples. (b) The enlarged view of the normalized Sr $K$-edge EXAFS spectra for the $x=$ 0 sample, which shows the KN$_1$ double-electron excitation feature and the resulting spectra after the manual removal of this feature. (c) The $k^3$ weighted $\chi (k)$ spectra and (d) its Fourier transformation for the $x=$ 0 sample before and after the subtraction of the KN$_1$ excitation feature, where arrows highlight the change in the corresponding features. (e) The ratio of the magnitude of the Sr--Sr/La and Sr--Co/Nb scattering paths as a function of $x$. (f--k) The fitting of the $|\chi(R)|$ and corresponding Im[$\chi(R)$] spectra in the window shown by the dotted green line, where the Im[$\chi(R)$] spectra in each panel are shifted downward for the clarity in the presentation. The asterisk symbol in (f) represent the peak arising from the KN$_1$ intra-atomic transition.} 
\label{DP_Sr_K}
\end{figure*}

To further interpret the behavior of main feature at $\approx$1.5~\AA $\space$ in the EXAFS spectra, i.e., the change in the strength of the Co--O and Nb--O scattering paths with the La substitution, we compare them with the respective white line intensity of Co and Nb $K$-edge XANES spectra in Figs.~\ref{DP_Co_K}(b) and (c), respectively. A strong correlation between the intensity of white line and degree of the octahedral distortion can be clearly observed, irrespective of the direction of octahedral distortion in two cases. In order to understand this we simulate the Nb $K$-edge XANES spectra for the $x=$ 0 sample by varying the extent of octahedral distortion in NbO$_6$ unit, as shown in Fig.~\ref{XANES_Nb_K3}(a). Here, only Nb--O$_a$ and Nb--O$_e$ scattering path distances were varied in order to systematically enhance the $\Delta_{\rm Nb}$, keeping all the other structural parameters constant. We find a systematic reduction in the intensity of the white line with increase in the octahedral distortion, as shown in the inset (a1) and plotted quantitatively in the inset (a2) of Fig.~\ref{XANES_Nb_K3}(a). The corresponding intensity of Nb $p$-LDOS also show the monotonic reduction with increase in the value of $\Delta_{\rm Nb}$ as indicated by the arrow in Fig.~\ref{XANES_Nb_K3}(b). The increase in the degree of the octahedral distortion is expected to enhance the orbital hybridization between Nb and oxygen states, and can result into the spatially wide distribution of the Nb $p$ states. The FEFF code simultaneously calculate the charge on the different valence orbitals and the resulting charge transfer from the initial neutral atoms \cite{Nesvizhskii_PRB_2000}, which can be useful to  understand the electronic structure of the underlined compounds and its evolution with the different perturbations, as presented in Table II for the selected atoms. Here, the $\Delta$e represent the charge transfer within the Norman radius at the time of the self-consistent calculations, which is the difference between the total $l$-projected charge count and number of valence electrons in the free atoms (considered one additional electron in the valence state of absorbing Nb, accounting for the core-hole) \cite{Modrow_PRB_03, Ankudinov_PRB_98}. Note that the $\Delta$e values are much smaller than the underlined oxidation state of Nb and O atoms in the sample, but these can be used to qualitatively understand the variation in the electronic structure with $\Delta_{\rm Nb}$. Also, the octahedral oxygen environment around the absorbing Nb atoms give rise to the $sp_3d_2$ type hybridization, causing a decrease in $s$ and increase in the $p$ and $d$ orbital electron count as compared to the free Nb and O atoms [$s=$ 1(2), $p=$ 6(4), $d=$ 4(0) for the Nb(O)] \cite{Modrow_PRB_03}. Importantly, the value of $\Delta$e decreases with increase in the octahedral distortion for both Nb as well as oxygen atoms, which indicate the enhancement in the degree of orbital overlapping between these two atoms, resulting in the deviation from the free ion like character of Nb atom. Further, an increase in the difference between Nb--O$_e$ and Nb--O$_a$ bond distances enhance the spatial distribution of Nb $p$ states (hybridized with the O 2$p$ states) and hence decrease the strength of the white line feature. This hypothesis suggest that the pure $1s\rightarrow np$ transition as an origin of the white line in the $K$-edge XANES spectra is an over simplified assumption and effect of the molecular orbitals on the XANES features of the transition elements should also be stressed. For example, the change in the XANES spectra of the transition metal oxides with varying degree of electronegativity have been well explained by Modrov $et$ $al.$ on the basis of the molecular orbital theory \cite{Modrow_PRB_03}.  

In Fig.~\ref{DP_Sr_K}(a) we present the magnitude of the Fourier transformed $k^3 \chi (k)$ of Sr $K$-edge spectra to investigate the local coordination environment around the $A$-site cations.  Interestingly, an intense peak is observed at a very small radial distance of around 0.8~\AA, which is smaller even than the ionic radii of Sr$^{2+}$ cation in the 12-fold coordination environment \cite{Shannon_AC_76} and hence no scattering path is expected at this position. These features at the low radial distances are usually not associated with the local structural arrangement of the atoms, but arises mainly due to the multielectronic transitions within the atoms, which can appear well above the white line in the absorption spectra and mix with the EXAFS oscillations \cite{ Li_PRL_92, Angelo_PRA_93, Filipponi_PRA_95, Angelo_PRB_96}. The inaccurate treatment of these spectral features can lead to the significant error in structural parameters extracted from the EXAFS analysis. In case of Sr $K$-edge absorption spectra, an intra atomic excitation from 1$s$ to 4$s$ (KN$_1$) states result in the sharp absorption feature well above the white line \cite{Angelo_PRB_96}. The position of this feature can be estimated by the Z+1 approximation, where binding energy of the Sr ($Z =$38) electrons in the different levels can be approximated by that of Yttrium ($Z=$ 39) after the ejection of one photoelectron from the $K$ shell. This approximation predict the KN$_1$ double electron excitation at 43.8~eV above the Sr $K$-edge threshold energy \cite{Angelo_PRB_96}. Interestingly a sufficiently intense feature is clearly observed at 44.4(4)~eV above the threshold energy of Sr $K$-edge in all the samples, as shown in Fig.~\ref{DP_Sr_K}(b) for the $x=$ 0 sample, see Fig.~14 of \cite{Kumar_XAS_SI}. The absence of this feature above the white line of the La L$_3$-edge spectra, as indicated by the black arrow in Fig.~\ref{XANES_Sr_K}(c) for the $x=$ 1 sample, clearly indicate the atomic-like states of the Sr atom as its origin rather than the local atomic arrangement around the Sr atoms. Therefore, a careful treatment of the Sr $K$-edge EXAFS spectra is necessary for the accurate structural probe. In order to further confirm the origin of this low radial distant peak in Fig.~\ref{DP_Sr_K}(a), we subtract the KN$_1$ excitation feature by a Gaussian approximation and the resulting spectrum is presented in Fig.~\ref{DP_Sr_K}(b) for the comparison. We find a clear elimination of a spectral feature around $k=$ 4 \AA$^{-1}$ as a result of the subtraction of KN$_1$ double-excitation transition, as highlighted by arrow in Fig.~\ref{DP_Sr_K}(c) and a well speculated sharp reduction at around 0.8~\AA $\space$ in the magnitude of Fourier transformed $k^3 \chi(k)$ spectra of Sr $K$-edge in Fig.~\ref{DP_Sr_K}(d). However, the intensity of few other features also changes as a result of this subtraction, possibly due to inaccuracy in the approximation of KN$_1$ feature in the spectra resulting from the presence of other nearby spectral features from the local atomic arrangement. Thus, we analyze the as recorded EXAFS spectra of Sr $K$-edge for all the samples by excluding the low radial part of the $|\chi(R)|$ to avoid the contribution from the KN$_1$ excitation channel. Also, the KM$_{2,3}$ and KM$_{4,5}$ excitations are possible in the Sr $K$-edge EXAFS spectra resulting from 1$s$ $\rightarrow$ 3$p$ and 1$s$ $\rightarrow$ 3$d$ transitions, respectively \cite{Angelo_PRB_96}. However, no change is observed in the spectra at their speculated positions from the $Z$+1 approximation in any of the samples due to dominance of the high amplitude EXAFS features resulting from the good micro crystalline nature of the samples. However, the effect of these transitions can not be completely neglected in the analysis of the Sr $K$-edge EXAFS spectra and the data up to very high temperature ($>$2000~K) are required to fairly quantify their contribution, where structural oscillations significantly diminish due to the enhanced DW factor \cite{Filipponi_PRA_95}. 

Moreover, we observe three intense features in the Sr $K$-edge EXAFS spectra between 1.4--3.5~\AA $\space$ for all the samples [see Fig.~\ref{DP_Sr_K} (a)]. These features at around 2, 2.6, and 3.2~\AA $\space$ result from the Sr--O, Sr--Co/Nb, and Sr--Sr/La scattering paths, respectively. Also, a monotonic reduction in the intensity of the peaks corresponding to Sr--Co/Nb and Sr--Sr/La scattering paths is clearly visible with the La substitution [see Fig.~\ref{DP_Sr_K}(a)]. However, a small inconsistency is present in case of the $x=$ 1 sample possibly due to decrease in the quality of the signal because of the lower Sr concentration. 
In $I4/m$ symmetry, two different oxygen positions lead to three sets of Sr--O bond distances with four bonds in each group, which further split into twelve different bonds in $P2_1/n$ symmetry \cite{Kumar_NPD_22}. However, we take three Sr--O scattering paths with the same DW factor to fit the Sr $K$-edge EXAFS spectra in 1.4--3.5 \AA $\space$ range for all the samples, as shown in Figs.~\ref{DP_Sr_K}(f--k) and best fit parameters are presented in Table I. Further, narrow spatial distribution unable us to distinguish the Sr--Co and Sr--Nb as well as Sr--Sr and Sr--La paths and we take single Sr--Co/Nb and Sr--Sr/La paths for the EXAFS fitting. A reduction in the amplitude of the Sr--Co/Nb and Sr--Sr/La peaks indicate the distortions in the corresponding scattering paths with the La substitution, which is reflected into their corresponding Debye-Waller factors, as presented in Table I. This effect is more prominent in case of Sr--Co/Nb as compared to Sr--Sr/La scattering path, which is clear in the plot of amplitude ratio of these two scattering paths [(Sr--Sr/La)/(Sr--Co/Nb)] particularly for the $x\geqslant$0.6 samples, see Fig.~\ref{DP_Sr_K}(e). The change in formal charge and ionic radii of $A$- (Sr$^{2+} \rightarrow$ La$^{3+}$) and $B$- (Co$^{3+} \rightarrow$ Co$^{2+}$) site atoms cause the reduction in amplitude of their respective scattering paths with $x$. Further, the different tendency of the Nb--Sr/La (increasing) and Co--Sr/La (decreasing) SS paths with $x$ is evident from the Nb and Co $K$-edge EXAFS spectra, respectively, results in the spatially wide distribution and hence more rapid reduction in the intensity of Sr--Co/Nb path as compared to Sr--Sr/La with $x$. Moreover, the two features between 4--5 \AA $\space$ are attributed to the SS from the second and third nearest oxygen atoms, located in the neighboring pseudocubic unit cells and a feature around 5.5 \AA $\space$ is resulting from the Sr--Sr2/La2 SS  path. A well speculated distortion in these scattering paths cause the observed reduction in the intensity of these features with $x$. This behavior suggests that the chemical pressure due to La substitution induce more distortion at the $A$-site as compared to the $B$-site(s), and results in the evolution of complex magnetic interactions \cite{Kumar_PRB1_20, Kumar_PRB2_20}. 

\begin{figure}[h]
\includegraphics[width=3.45in]{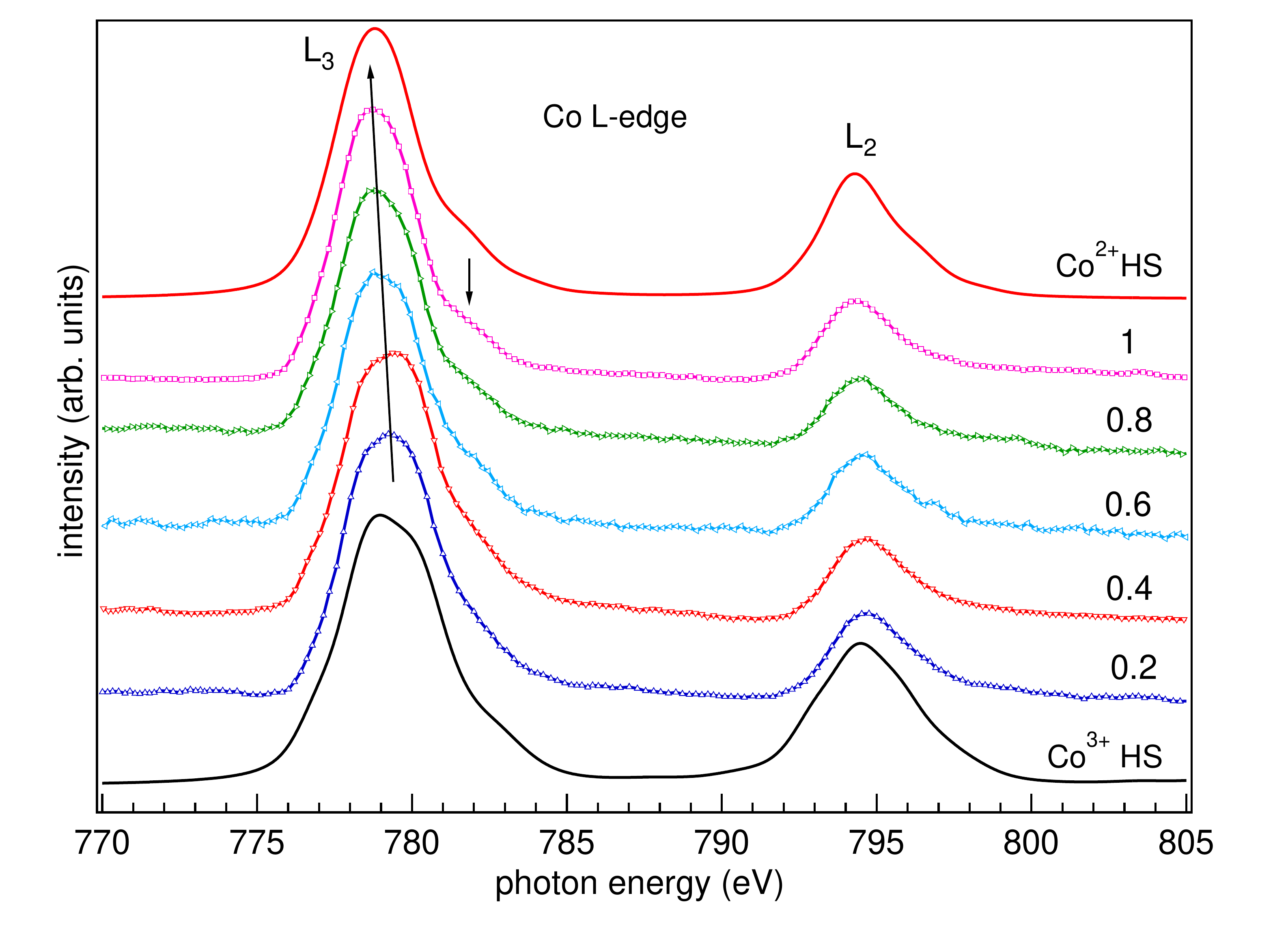}
\caption {The Co L$_{2,3}$-edge spectra (vertically offset for clarity) for Sr$_{2-x}$La$_x$CoNbO$_6$ samples. The atomic multiplet based simulated spectra for HS Co$^{2+}$ and HS Co$^{3+}$ are compared.} 
\label{SXAS}
\end{figure}

Finally, we present the Co L$_{2,3}$ soft-XAS spectra in Fig.~\ref{SXAS} where transition from two spin-orbit splitting components of Co $2p$ final states, i.e., 2$p_{3/2}\rightarrow 3d$ (L$_3$) and 2$p_{1/2}\rightarrow 3d$ (L$_2$), are observed around 780~eV and 795~eV, respectively, for all the samples. The peak position shifts towards lower energy (as indicated by the arrow) due to the conversion of Co from 3+ to 2+ valence state with the La substitution. This is also consistent with the Co $K$-edge XANES spectra, discussed above. Interestingly, we observe the consistent reduction in the broadening with the evolution of a hump at the shoulder of L$_3$ edge with the La substitution, as indicted by the vertical arrow for the $x=$ 1 sample. To understand the physical origin of this, we simulate the $L$-edge XAS spectra for HS Co$^{2+}$ and HS Co$^{3+}$ using the CTM4XAS software. A charge transfer effect from oxygen to Co atoms was also introduced in the simulation for both the valence states of Co, resulting in the additional $2p^63d^7\underline L^1 \rightarrow 2p^53d^8 \underline L^1$ and $2p^63d^8\underline L^1 \rightarrow 2p^53d^9 \underline L^1$ transitions in case of Co$^{3+}$ and Co$^{2+}$, respectively. The Lorentzian and Gaussian line shapes with half-width of 0.5~eV and 0.45~eV were used to account for the core-hole lifetime and instrumental broadening. The crystal field-field splitting (10Dq), on-site Coulomb correlation energy of 3$d$ electrons ($U_{dd}$), core hole potential ($U_{pd}$), charge transfer energy ($\Delta_{ct}$), and O 2$p$-Co 3$d$ hybridization strength parameters were varied to best match the experimental and simulated spectra. The intra-atomic interactions were approximated by reducing the Slater integrals to 80\% from their Hartree-Fock values. The best matched parameters are 10Dq=1.3 and 0.5~eV, $U_{dd}-U_{pd}$ =3.0 and 0.2~eV, and $\Delta_{ct}$=--6.5 and 1.4~eV for HS Co$^{3+}$ and HS Co$^{2+}$ spectra, respectively. The $p-d$ hybridization strength parameter T$_{e_g}$=2.0 and 0.8, and T$_{t_{2g}}$=1.0 and 0.4 were used for Co$^{3+}$ and Co$^{2+}$, respectively and the resulting spectra are presented in Fig.~\ref{SXAS}. The extracted value of 10Dq for Co$^{3+}$ is smaller than that of LaCoO$_{3}$ ($\sim$2.3~eV \cite{Medarde_PRB_06}), indicating the higher probability of finding Co$^{3+}$ in the IS/HS state, which is consistent with the magnetization results \cite{Kumar_PRB1_20, Yoshii_JalCom_2000}. Further, the large negative value of the charge transfer energy indicate the highly covalent nature of the Co--O bonds, where the covalency character decreases with the La substitution. Moreover, the branching ratio, defined as (BR)= I(L$_3$)/[I(L$_2$)+I(L$_3$)], where I(L$_2$) and I(L$_3$) represent the integrated intensity of L$_2$ and L$_3$ edges, is 0.757, 0.786, 0.764, 0.763, and 0.758 for the $x=$ 0.2--1 samples, respectively. These preliminary results of SXAS further suggest for the mixed IS and HS states of Co$^{3+}$ in Sr$_{2-x}$La$_x$CoNbO$_6$  samples \cite{Thole_PRB_88}. However, high resolution SXAS measurements can be further useful to understand the complex spin-state transition in these samples as a function of $x$ and temperature \cite{Kumar_PRB1_20, Merz_PRB_11}. 

\section{\noindent ~Conclusions}

We have investigated the electronic and local structure of Sr$_{2-x}$La$_x$CoNbO$_6$ ($x=$ 0--1) samples using detailed analysis of XANES and EXAFS spectra as well as the simulation of Nb, Co and Sr $K$- edges. A very weak pre-edge feature in case of Nb $K$-edge XANES spectra indicate the negligible off-center displacement of Nb atoms in the NbO$_6$ octahedra. This pre-edge feature results from the transition of 1$s$ electrons to the 4$d$ states of Nb mixed with Nb and/or O $p$ states. Further, the post edge feature in the Nb $K$-edge is found to vary sensitively with both nature as well as distance of the $A$-site atoms. The $K$-edge XANES spectrum show that Co transform from 3+ to 2+ valence states, whereas oxidation states of Sr and Nb remains invariant with the La substitution. Also, the intensity of pre-edge features of Co $K$-edge spectra depends on the number of 3$d$ electrons, but analysis discard their pure quadruple origin. Moreover, the significantly strong pre-edge feature in case of Co $K$-edge XANES spectra as compared to Nb indicate the presence of Co and Nb atoms in the different octahedral environment. The EXAFS measurements evident a decrease in the local distortion around $B$ site atoms, despite of reduction in the overall crystal symmetry with the La substitution. The $z$-out JT distortion in CoO$_6$ and consequently $z$-in distortion in the NbO$_6$ octahedra have been observed in all the samples. The distortion in the CoO$_6$ octahedra decreases monotonically with the La substitution due to  increase in the concentration of JT inactive Co$^{2+}$ ions, whereas that in NbO$_6$ octahedra remains almost invariant up to the $x=$ 0.4, and then decreases for higher La concentration. This abrupt change in the local coordination environment around the Nb atoms results in the sharp enhancement in B-site ordering and hence evolution of AFM interactions in the $x\geqslant$0.6 samples. This is further supported by sudden enhancement in the longer Nb--Co2/Nb2 path for the $x\geqslant$0.6 samples in the higher coordination shell in Nb $K$-edge EXAFS spectra. Interestingly, we observe a strong correlation between the degree of octahedral distortion in (Co/Nb)O$_6$ units and intensity of their respective white line features. An increase in the degree of octahedral distortion is found to enhance the metal-ligand overlapping and delocalized nature of the TM $p$-states, resulting in the low intense broad white line features in their $K$-edge XANES spectra. We reveal that such a correlation is useful to estimate the extent of octahedral distortion in the complex compounds directly from the analysis of XANES spectra. Moreover, the Sr $K$-edge absorption spectra show clear evidence of the KN$_1$ double electron excitation for all the samples, suggesting the necessity of the careful treatment of the Sr $K$-edge EXAFS spectra. Finally, the Co $L_{2,3}$ edge spectra indicate the decrease in the crystal field strength and tendency of O to Co charge transfer with the La substitution ($x$). \\ 

\section{\noindent ~Acknowledgments}

AK and RS thank the UGC and DST-Inspire for the fellowship. We thank John J. Rehr for very useful discussions on the XANES simulations using FEFF code. Authors also acknowledge D. M. Phase and R. K. Sah for support and help, respectively, during the soft XAS measurements. We acknowledge the financial support from  SERB-DST through core research grant (project reference no. CRG/2020/003436). We also greatly acknowledge the CRS (project no. CSR-ICISUM-36/CRS-319/2019-20/1371) for collaborative research work.

\end{document}